\documentclass[12pt]{iopart}
\usepackage{times}
\usepackage{color}
\usepackage[colorlinks,bookmarks=false,citecolor=blue,linkcolor=red,urlcolor=blue]{hyperref}
\usepackage{iopams}
\usepackage{colortbl,amsthm,amssymb,txfonts}
\usepackage{graphicx}
\usepackage{epsfig,color}

\begin{document}
 
\title{Pauli limited d-wave superconductors: Quantum breached pair phase and thermal transitions}
\author{Madhuparna Karmakar}
\address{Indian Institute of Technology, Madras, Chennai-600036, India.}
\ead{madhuparna.k@gmail.com}
\begin{abstract}

We report a quantum phase transition in Pauli limited $d$-wave  
superconductors and give the mean field estimates of the associated quantum critical point. 
For a population imbalanced $d$-wave superconductor a stable ground state phase viz. quantum 
breached pair phase has been identified
which comprises of spatial coexistence of 
gapless superconductivity and nonzero magnetization. Based on the thermodynamic 
and quasiparticle indicators we for the first time analyze this phase, 
discuss the thermal behavior of Pauli limited $d$-wave superconductor, give accurate 
estimates of the thermal scales associated with such systems and map out the pseudogap regime. 
Our work shows that while the Pauli limited superconductors are known to exhibit exotic 
modulated superconducting phase at large imbalance of fermion populations; 
in the regime of weak imbalance an intriguing phase of competing orders is realized.
We have established that rather than the superconducting pairing field, it is the average 
magnetization of the system that quantifies this quantum phase transition.
Given that the existing Pauli limited superconductors possess unconventional pairing 
state symmetry of the superconducting order, our work promises to open up new avenues in 
the experimental research of these materials. We have also demonstrated an alternate
scenario wherein the quantum breached pair phase is a natural outcome for a $d$-wave 
superconductor with unequal effective masses of the fermion species.   
\end{abstract}

\date{\today}

\section{Introduction}

Superconductivity, in competition or coexistence with magnetic correlations has been a 
primary area of research in condensed matter physics, over the past few 
decades. 
While magnetism is usually considered to be detrimental towards the existence of 
superconducting order, there are examples of materials where 
magnetic correlations reside proximate to or in coexistence with conventional 
\cite{fulde1982,muller2001,lynn1997,braun2001,schneider2009,baba2008,schultz2011} 
or unconventional \cite{lee_rmp,pfleiderer_rmp,stewart_rmp,steglich2007,fisk2006,johnson2010,wu2008,fang2008,chen2010,yuan2011} superconducting orders. The fascinating phenomena of coexisting orders 
as observed in these materials is dictated by the precise tuning of the 
energy landscape, often carried out via external control parameters such as, doping, 
pressure etc \cite{lee_rmp,pfleiderer_rmp,stewart_rmp,dagotto2012,johrendt2011}. 
\begin{figure}
\begin{center}  
\includegraphics[height=7.5cm,width=7.5cm,angle=0]{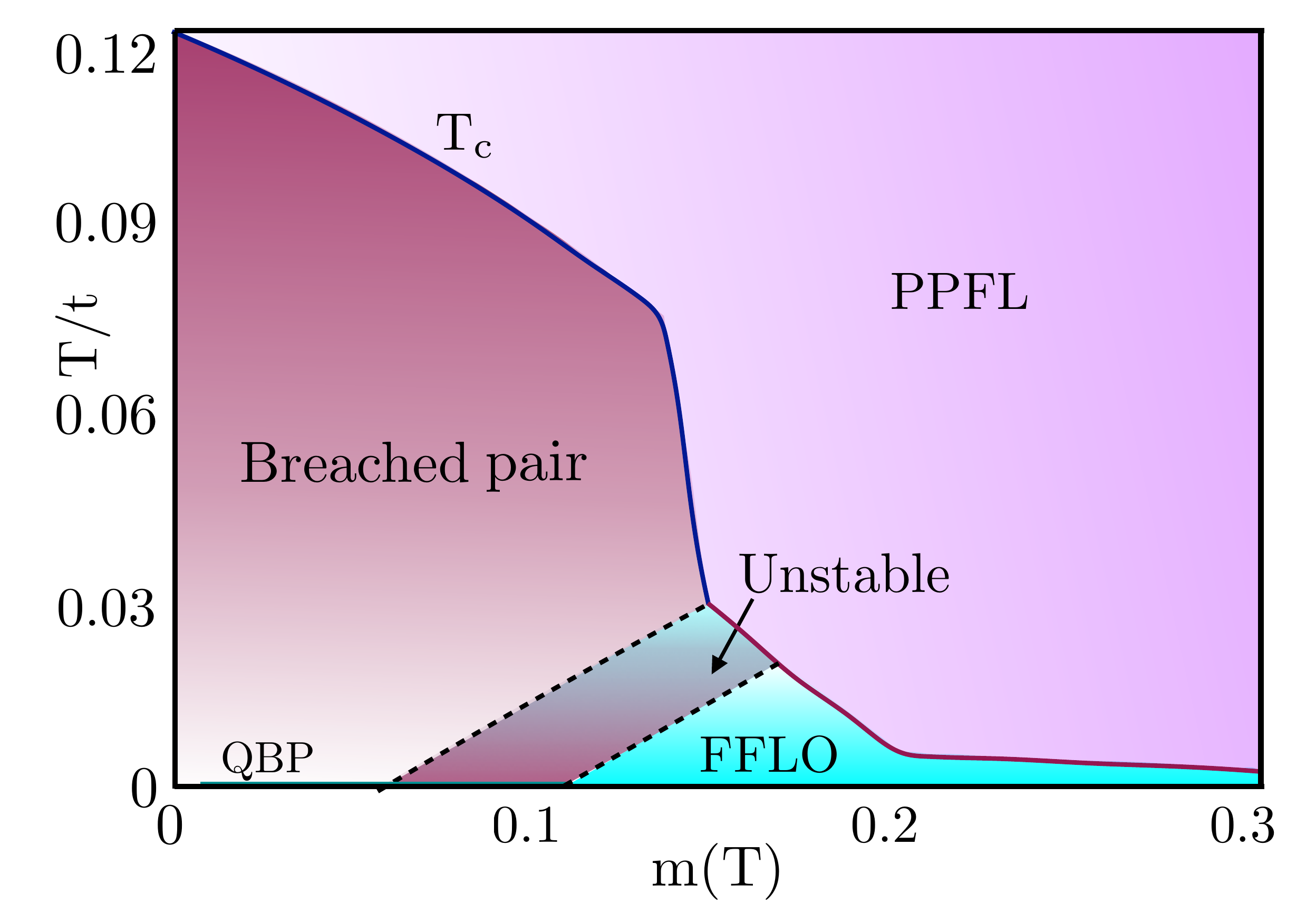}
\caption{Color online: Magnetization-temperature ($m-T$) phase diagram of population imbalanced 
$d$-wave superconductor. The solid curve correspond to $T_{c}$, the low $m$ regime of second order 
transition is shown by the black curve while the red curve at high $m$ correspond to the regime of first 
order phase transition.}
\end{center}
\label{finite_pd_mt}
\end{figure}

An equally intriguing but relatively less explored phenomenon of  
coexisting superconducting and magnetic correlations is observed in Pauli limited superconductors. 
The Pauli limited superconductors are characterized by their loss of superconducting order via 
Pauli paramagnetic pair breaking effect, which originates from the Zeeman splitting of the single 
electron energy levels \cite{shimahara2007,saintjames}. When a superconducting pair undergoes 
Pauli paramagnetic pair breaking it gives rise to unpaired fermions in the system. In the presence 
of such unpaired fermions an 
uniform (zero momentum paired) superconducting state is no longer stable, rather the 
superconducting pair acquires a finite momentum which shows up as spatial modulation in the 
superconducting order. The unpaired fermions coexist with such spatially 
modulated superconducting state giving rise to what is now well known as the Fulde-Ferrell-Larkin-Ovchinnikov 
(FFLO) superconductivity \cite{ff1964,lo1964}. Pauli limited superconductivity has been realized 
both in solid state (e. g. CeCoIn$_{5}$ \cite{bianchi2003,tayama2002,koutroulakis2010,kenzelmann2008,capan2004,martin2005,kenzelmann2014,matsuda2006,kenzelmann_prx2016,movsovich2019},
$\kappa$-BEDT \cite{wosnitza2013,lortz2007,mitrovic2014,wright2011,coniglio2011,lortz2011,agosta2012,cho2009},
KFe$_{2}$As$_{2}$ \cite{zocco2013,cho2011,khim2011} etc.) as well as in ultracold atomic gas 
\cite{ketterle2008,ketterle2007,ketterle2006,liao2010} systems. 
The central requirement for realizing Pauli limited superconductivity is the creation of an imbalance 
in the population of the fermionic species and thus a Fermi surface mismatch, 
via an applied Zeeman field in solid state systems or by 
loading different populations of the fermionic species in optical lattice in ultracold atomic 
gas set ups.     

While much attention has been paid to the FFLO phase over the past few years
\cite{bianchi2003,tayama2002,koutroulakis2010,kenzelmann2008,capan2004,martin2005,kenzelmann2014,matsuda2006,kenzelmann_prx2016,movsovich2019,wosnitza2013,lortz2007,mitrovic2014,wright2011,coniglio2011,lortz2011,agosta2012,cho2009,zocco2013,cho2011,khim2011}, 
another exotic phase of the Pauli limited superconductors exhibiting spatial coexistence of superconductivity and non zero 
magnetization, remains almost unexplored. The said phase is known as the ``Breached pair'' (BP) and 
corresponds to a situation where the imbalance in fermionic population is not strong enough to give 
rise to a FFLO phase but is sufficient to give rise to a Fermi surface mismatch \cite{mpk2016,mpk_epj2016,torma_njp}. 
The issue of the BP phase was first raised decades ago by Sarma \cite{sarma1963} in his seminal work,  where 
he discussed the possibility of self consistent mean field solution with gapless mode,  in $s$-wave superconductors,  
in presence of an applied magnetic field. However, he found that such a phase if it exists would be 
energetically unfavorable as compared to the uniform superfluid (BCS) phase. 

\begin{figure}
\begin{center}
\includegraphics[height=12cm,width=14cm,angle=0]{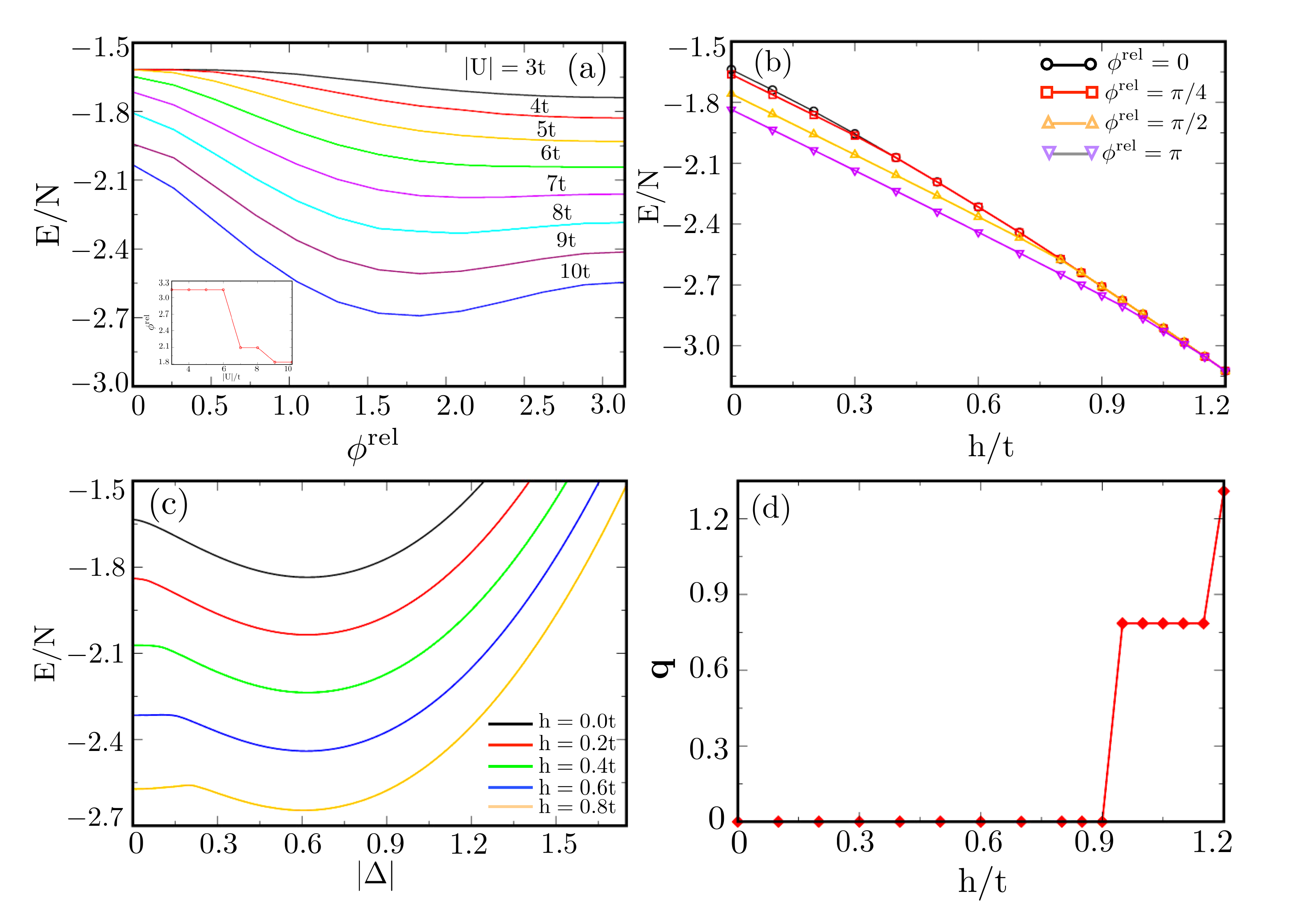}
\caption{Color online: Ground state energy landscape. (a) Energy optimization w. r. t. $\phi^{rel}$ 
at different $\mid U\mid/t$.  
Inset: $\phi^{rel}$ corresponding to minimum energy state at different $\mid U\mid/t$. 
(b) Energy w. r. t population imbalance (Zeeman field) at different $\phi^{rel}$ for $\mid U\mid = 4t$ 
and $\mu=-0.2t$. The lowest energy state correspond to $d_{x^{2}-y^{2}}$ 
pairing state symmetry with $\phi^{rel}=\pi$. (c) Optimization of pairing field amplitude $\mid \Delta\mid$ 
at different $h/t$ for $\mid U\mid =4t$, $\mu=-0.2t$ and $\phi^{rel}=\pi$. 
(d) Optimized pairing momentum (${\bf q}$) at different Zeeman field ($h/t$) at $\mid U\mid = 4t$, 
$\mu=-0.2t$ and $\phi^{rel}=\pi$.}
\end{center}
\end{figure}
Advent of ultracold atomic gas experiments brought renewed interest in the physics of the BP phase. 
Through detailed analytic calculations it was demonstrated by Liu and Wilczek \cite{wilzek, wilzek_prl,wilzek2003} 
that while for an imbalanced Fermi system the usual single band
Hubbard model with on site interactions between the fermion species does not allow the BP phase to be 
a stable ground state, the same is indeed possible under non trivial circumstances 
such as, (i) imbalance in fermion effective masses in presence of contact interaction between them, 
(ii) momentum dependent interaction, (iii) same species repulsion etc. \cite{wilzek, wilzek_prl,wilzek2003}.
It was further shown that for suitable choice of parameters a mass imbalanced $s$-wave superconductor 
hosts the BP phase as a stable ground state,  which undergoes a second order quantum phase transition to the 
uniform superconducting phase as a function of decreasing Zeeman field \cite{wilzek}. 
This transition was said to belong to the class of {\it topological Lifshitz transitions} in the sense that 
the Fermi surface topology changes across this transition \cite{lifshitz}. Note that this topological 
transition does not involve the change in any topological invariant.
The fate of the BP phase in an imbalanced unconventional (non $s$-wave) superconductor however continues 
to remain an open question till date, in spite of almost all the known Pauli limited superconductors being
 unconventional in their pairing state symmetry. 
\begin{figure}
\begin{center}
\includegraphics[height=7.5cm,width=12.5cm,angle=0]{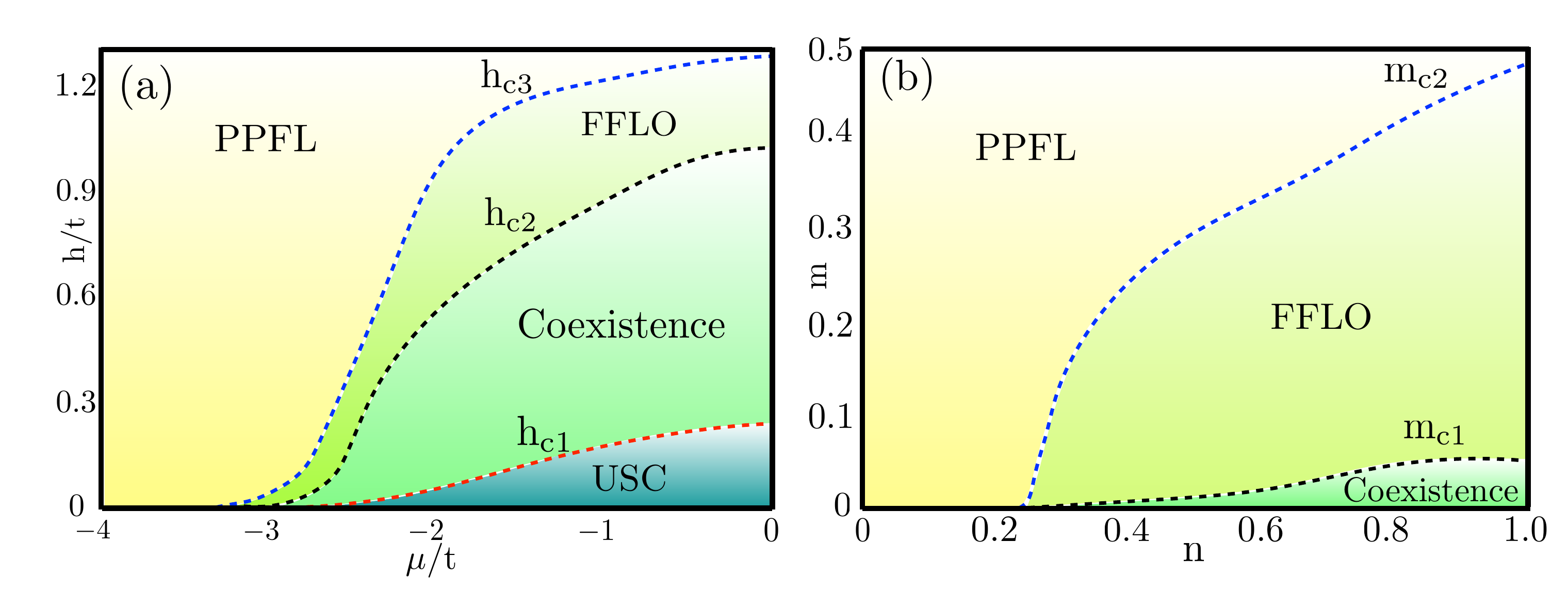}
\caption{Color online: Ground state phase diagram showing the 
thermodynamic phases in the (a) chemical potential-Zeeman field ($\mu-h$) and (b) number density-
magnetization ($n-m$) plane. The order of phase transition from USC to coexistence, from coexistence to FFLO and 
from FFLO to PPFL are second, first and second, respectively, and are marked by the corresponding 
critical fields and magnetization. In the $n-m$ plane the USC phase collapses to the x-axis.}
\end{center}
\label{gs_pd_muh}
\end{figure}

A sufficient volume of literature has already been devoted to theoretically
capture the physics of FFLO phase in the strong imbalance regime of Pauli limited nodal ($d$-wave) superconductors 
\cite{graf2005,graf2006,spalek2011,vekhter2010,yanase2008,graf2008,ting2009,kunyang}. 
In the present paper we focus on the regime of weak population imbalance in such superconductors  
and for the first time demonstrate that the BP phase is a stable ground state over a significant regime 
of population imbalance in Pauli limited $d$-wave superconductors.
We refer this ground state BP phase as the ``Quantum Breached Pair'' (QBP) phase 
and establish that this phase comprises of zero-momentum {\it gapless} superconductivity and 
nonzero magnetization, in coexistence. A second order quantum phase transition takes place between the QBP 
and unmagnetized $d$-wave superconductor (USC) phases, however, there is no explicit symmetry 
breaking across this phase transition.
We emphasize that the superconducting pairing field is not a suitable order 
parameter when it comes to quantifying the phase transition between QBP and USC phases, 
as the pairing field remains unchanged across this transition. The phase transition between the 
QBP and USC phases 
is rather characterized by an alternative order parameter viz. magnetization ($m_{i}$).
As we demonstrate in the following sections, the ground state phases of the 
Pauli limited $d$-wave superconductor are categorized based on the,  (i) superconducting 
pairing field ($\Delta_{ij}$), (ii) single particle density of states at the Fermi level ($N(0)$) and 
(iii) magnetization ($m_{i}$) as, (a) USC ($\Delta_{ij}({\bf q}=0) \neq 0$, $N(0)=0$, $m_{i} = 0$), 
(b) QBP ($\Delta_{ij}({\bf q}=0) \neq 0$, $N(0) \neq 0$, $m_{i} \neq 0$), (c) FFLO 
($\Delta_{ij}({\bf q}\neq 0)\neq 0$, $N(0) \neq 0$, $m_{i} \neq 0$) and 
(d) partially polarized Fermi liquid (PPFL) ($\Delta_{ij} = 0$, $N(0) \neq 0$, 
$m_{i} \neq 0$), where, ${\bf q}$ corresponds to the superconducting pairing momentum. 
\begin{figure*}
\begin{center}
\includegraphics[height=4.5cm,width=16cm,angle=0]{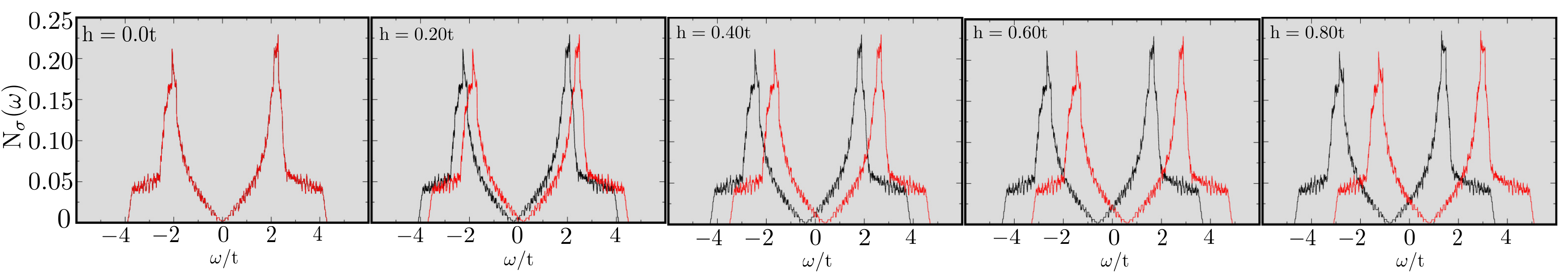}
\caption{Color online: Zeeman field dependence of spin resolved  fermionic
density of states ($N_{\sigma}(\omega)$), at $T=0$.
The black (red) curves correspond to $N_{\uparrow}(\omega)$
($N_{\downarrow}(\omega)$), respectively. Note the accumulation of spectral weight at the
Fermi level ($\omega=0$) with increasing field.}
\end{center}
\label{gs_dos}
\end{figure*}

Our real space approach to the problem enables 
us to identify and map out spatial coexistence of gapless $d$-wave superconducting pairing and nonzero
magnetization, characteristic to the QBP phase. Apart from identifying the phase transitions at the
ground state, the merit of this work rests in tracking the thermal
evolution of these phases using a non perturbative numerical technique, mapping out the relevant thermal scales
associated with the system and demonstrating that the QBP phase smoothly crosses over to its 
classical counterpart with increasing temperature. It must be noted that within the premises of lattice 
fermion models the existing works addressing the thermal physics of the Pauli limited $d$-wave superconductors 
are limited to mean field theory (MFT)  \cite{ting2009,kunyang} which are known to significantly overestimate 
the thermal scales in the interaction regimes away from the weak coupling limit \cite{randeria_taylor, mpk2016}.
 The unconventional $d$-wave superconductors are established to be in the regime of intermediate coupling, which 
renders the MFT unsuitable to address the thermal behavior of these systems. 
Our technique takes into account the thermal (spatial) fluctuations of the pairing field 
{\it at all orders} and not just the saddle point fluctuations. Consequently, it is well suited to capture the 
thermal scales associated with the system accurately and gets progressively accurate with increasing 
temperature. As $T \rightarrow 0$, the thermal fluctuations die out and our approach becomes akin to the 
MFT at the ground state. In this spirit the quantum phase transition discussed in this work is basically 
the mean field estimate of the same. While we do not expect any qualitative change in our ground state results 
via inclusion of quantum fluctuations, quantitative shifts in the ground state phase boundaries are possible.

While we discuss our results in the following sections, we highlight our principal observations here: 
(a) We for the first time demonstrate that the breached pair phase is a stable ground state of 
the Pauli limited $d$-wave superconductor. We refer to this ground state breached pair phase as QBP 
and show that it comprises of spatial coexistence of zero-momentum gapless superconducting 
order and nonzero magnetization. 
(b) By tuning the applied Zeeman field (or imbalance in the populations of the fermionic species) a second
 order quantum 
phase transition can be realized between the USC and QBP phases, 
quantified by the evolution of magnetization. In this work,  we give the mean field 
estimate of this quantum phase transition. (c) Based on the 
thermodynamic and quasiparticle signatures we track the thermal evolution 
of the ground state phases and map out the relevant thermal scales; we demonstrate that short range 
superconducting pair correlations survive up to temperatures $T \gg T_{c}$ and gives rise to the 
``pseudogap'' regime. (d) We 
show that for a $d$-wave superconductor with imbalance in the effective masses of the fermion species, 
a QBP phase is realized in the absence of any imbalance in population. 
The thermal phase diagram of such a system reveals the existence of ``species dependent'' pseudogap scales 
which should be experimentally detectable.

The rest of the paper is organized as follows. In section II we discuss the model and the numerical 
technique used in this work, as well as the indicators used to characterize the phases, section III 
comprises of the main results of this work and their analysis. We discuss the mass imbalanced 
$d$-wave superconductors in section IV,  and follow it up by drawing the conclusions of our work. 

\section{Model, method and indicators}

In this section we sketch out the steps involved in studying the many body quantum Hamiltonian, 
discuss our choice of the parameters and the relevant indicators to characterize the phases of this 
system.

\subsection{Model}

Our starting Hamiltonian corresponds to a two-dimensional (2D) square lattice
with nearest neighbor attractive interaction between the fermions, in presence of a Zeeman field,
and reads as,
\begin{eqnarray}
H & = & -\sum_{\langle ij\rangle, \sigma}t_{ij}c^{\dagger}_{i,\sigma}c_{j,\sigma}-U\sum_{\langle ij\rangle}\hat n_{i} \hat n_{j}
 -\mu\sum_{i,\sigma}\hat n_{i,\sigma} - h\sum_{i, \sigma}\sigma_{i}^{z}\hat n_{i,\sigma}
\end{eqnarray}

\noindent where, $t_{ij}$ corresponds to the hopping parameter such that $t_{ij}=t=1$
for the nearest neighbors and is zero otherwise, $\mid U \mid > 0$ corresponds to the nearest neighbor attractive
interaction between the fermions, the net number density of the fermions in the system is maintained
through the chemical potential $\mu =(1/2)(\mu_{\uparrow}+\mu_{\downarrow})$, 
while $h=(1/2)(\mu_{\uparrow}-\mu_{\downarrow})$ corresponds to the effective Zeeman field. 

\subsection{Method}
The partition function is basically a functional integral over the 
Grassman fields $\psi_{i\sigma}(\tau)$ and $\bar\psi_{i\sigma}(\tau)$

\begin{eqnarray}
Z  =  \int {\cal D} \psi {\cal D} {\bar \psi}e^{-\int^{\beta}_{0} d\tau {\cal L}(\tau)} \cr
{\cal L}(\tau) = {\cal L}_{0}(\tau) + {\cal L}_{U}(\tau) \cr 
{\cal L}_{0}(\tau)=\sum_{\langle ij\rangle, \sigma}\{\bar \psi_{i\sigma}((\partial_{\tau}-\mu)\delta_{ij}+t_{ij})\psi_{j\sigma}\} \cr 
{\cal L}_{U}(\tau) = -U\sum_{\langle ij\rangle \sigma \sigma^{\prime}}\bar \psi_{i\sigma}\psi_{i\sigma}\bar\psi_{j\sigma^{\prime}}\psi_{j\sigma^{\prime}}
\end{eqnarray}

\noindent where, $\beta$ is the inverse temperature. Our strategy is to decompose $\hat n_{i}\hat n_{j}$ in terms of the bosonic auxiliary $d$-wave pairing singlet $\Delta_{ij}(\tau)$ using Hubbard-Stratonovich transformation \cite{hs,hs1}.
$ij$ and $\tau$ refers to the spatial and imaginary time dependence of the pairing field, respectively. In 
terms of the Matsubara frequency $\Omega_{n}=2\pi nT$ the pairing field reads as, $\Delta_{ijn}$, where $T$ 
is temperature. This leads to,

\begin{eqnarray}
Z = \int {\cal D}\psi {\cal D} {\bar \psi} {\cal D}\Delta {\cal D}{\Delta}^{*}e^{-\int^{\beta}_{0} d\tau {\cal L}(\tau)} \cr 
{\cal L}(\tau) = {\cal L}_{0}(\tau) + {\cal L}_{U}(\tau) + {\cal L}_{cl}(\tau) \cr 
{\cal L}_{0}(\tau)=\sum_{\langle ij\rangle, \sigma}\{\bar \psi_{i\sigma}((\partial_{\tau}-\mu)\delta_{ij}+t_{ij})\psi_{j\sigma}\} \cr
{\cal L}_{U}(\tau) = -\sum_{i \neq j} \Delta_{ij}(\bar \psi_{i\uparrow} \bar \psi_{j\downarrow}+\bar \psi_{j\uparrow} \bar\psi_{i\downarrow}) + h. c. \cr
{\cal L}_{cl}(\tau) = 4\sum_{i \neq j}\frac{\mid \Delta_{ij}\mid^{2}}{\mid U\mid}
\end{eqnarray}

Since the fermions are now quadratic, the $\int {\cal D}\psi {\cal D}{\bar \psi}$ integral can be performed to 
generate the effective action for the random background fields,

\begin{eqnarray}
Z = \int {\cal D}\Delta{\cal D}{\Delta}^{*} e^{-S_{eff}\{\Delta,\Delta^{*}\}} \cr
S_{eff} = \ln Det [{\cal G}^{-1}\{\Delta,\Delta^{*}\}] + \int^{\beta}_{0}d\tau{\cal L}_{cl}(\tau)
\end{eqnarray}

where, ${\cal G}$ is the electronic Green's function in the $\{\Delta\}$ background.
There are several options now, (i) Quantum Monte Carlo retains the full ``$i,\Omega_n$'' dependence of $\Delta$
computing $\ln [Det{\cal G}^{-1}\{\Delta\}]$ iteratively for importance
sampling. The approach is valid at all $T$, but does not readily yield
real frequency spectra. (ii) Mean field theory (MFT) is time independent, neglects the
phase fluctuations completely but can handle spatial inhomogeneity
in amplitude of the pairing field. Thus, $\Delta_{i}(i\Omega_n) \rightarrow \mid\Delta_{i}\mid$.
When the mean field order parameter vanishes at high temperature the theory trivializes.
(iii) Dynamical mean field theory (DMFT) retains the full dynamics
but keeps $\Delta$ at  effectively one site, {\it i.e},
$\Delta_{i}(\Omega_n) \rightarrow \Delta(\Omega_n)$.
This is exact when dimensionality $D \rightarrow \infty$.
(iv) Static path approximation (SPA) approach retains the full spatial dependence
in $\Delta$ but keeps only the $\Omega_n=0$ mode,
{\it i.e}, $\Delta_{i}(\Omega_n) \rightarrow \Delta_{i}$.
It thus includes classical fluctuations of arbitrary magnitude but no quantum ($\Omega_n \neq 0$)
fluctuations.  One may consider different 
temperature regimes. (1) $T = 0$: Since classical fluctuations die off at $T = 0$, SPA reduces to 
standard Bogoliubov-de-Gennes (BdG) MFT. 
(2) At $T\neq 0$ we consider not just the saddle-point configuration but {\it all configurations}.
These involve the classical amplitude and phase fluctuations of the order parameter, and the BdG equations 
are solved in {\it all these configurations} to compute the thermally averaged properties. 
This approach suppresses the order much quicker than in MFT. (3) High T : Since the $\Omega_{n}=0$ 
mode dominates the exact partition function, the SPA approach becomes exact as $T\rightarrow \infty$.
Consequently, it is akin to the MFT {\it only} at $T=0$ but captures the thermal physics 
of the system accurately.

We choose the last option (SPA) as our numerical technique. The resulting superconducting Hamiltonian reads as, 
\begin{eqnarray}
H_{SC} & = & -\sum_{\langle ij\rangle, \sigma}t_{ij}(c_{i\sigma}^{\dagger}c_{j\sigma} + h.c)
+ \sum_{i \neq j}\Delta_{ij} (c_{i\uparrow}^{\dagger}c_{j\downarrow}^{\dagger} +
c_{j\uparrow}^{\dagger}c_{i\downarrow}^{\dagger}) \nonumber \\ && + h. c. -\mu\sum_{i, \sigma} \hat n_{i\sigma}
- h\sum_{i, \sigma}\sigma_{i}^{z}\hat n_{i,\sigma} + 4\sum_{i \neq j}\frac{\mid \Delta_{ij}\mid^{2}}{\mid U \mid}
\end{eqnarray}
where, the last term of $H_{SC}$ corresponds to the classical stiffness
cost associated with the auxiliary field. Here, the $d$-wave singlet is defined as 
$\Delta_{ij} = (c_{i\uparrow}c_{j\downarrow}+c_{j\uparrow}c_{i\downarrow})$. It can further be expressed as
$\Delta_{ij} = \mid \Delta_{ij}\mid e^{i{\bf \phi}_{ij}}$, where ${\bf \phi}_{ij} \in \{\phi_{ij}^{x}, \phi_{ij}^{y}\}$
is the direction dependent phase of the complex pairing field and $\mid \Delta_{ij}\mid$ is the pairing field
amplitude, considered to be isotropic in the $xy$-plane. Note that in the usual mean field approach to $H_{SC}$ the
superconducting gap function $\Delta_{ij}$ is assumed to be a real number, but here we retain the degrees of freedom
associated with the pairing field phases and amplitudes. The system can thus be envisioned as free fermions 
moving on a random background of $\Delta_{ij}$.

The background field $\Delta_{ij}$ obeys the Boltzmann distribution,
\begin{eqnarray}
P\{\Delta_{ij}\} \propto Tr_{cc^{\dagger}}e^{-\beta H_{SC}}
\end{eqnarray}

\noindent which is connected to the free energy of the system. For large and 
random $\Delta_{ij}$ the trace is taken numerically. We generate the random 
background of $\{\Delta_{ij}\}$ by using Monte Carlo, diagonalizing $H_{SC}$ for each attempted 
update of $\Delta_{ij}$. The relevant fermionic correlators are computed
on the optimized configurations at different temperatures.
Evidently, the technique is computationally expensive. The computation
cost is cut down by using a traveling cluster approximation (TCA), wherein instead of diagonalizing $H_{SC}$ for
each attempted update of the auxiliary field, we diagonalize a smaller cluster surrounding the update site.
Both SPA and TCA have been extensively bench marked and used for several quantum many body systems.
\cite{tarat_epjb,mpk2016,nyayabanta_pyr,anamitra_prl}.

\begin{figure}
\begin{center}
\includegraphics[height=8cm,width=8cm,angle=0]{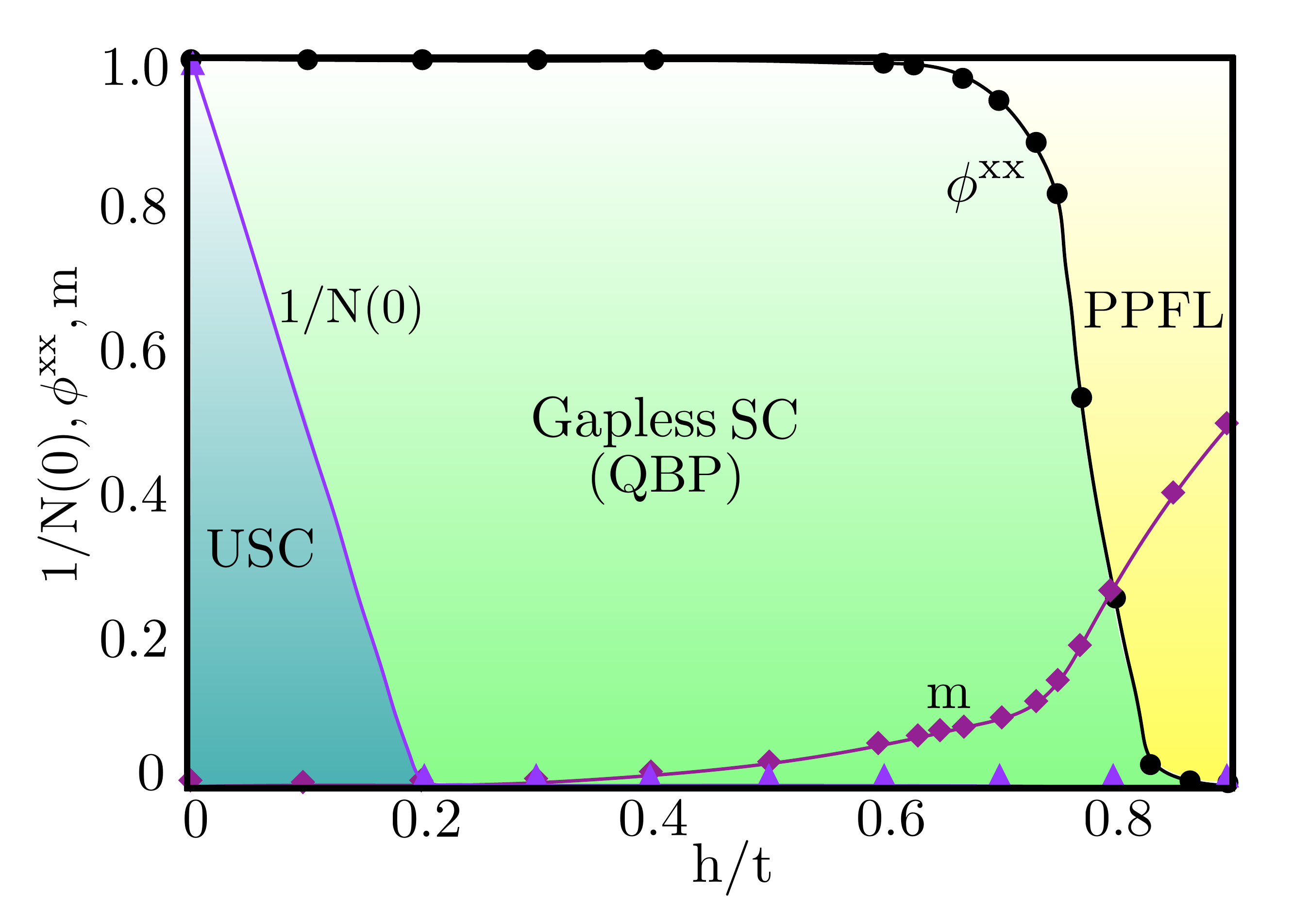}
\caption{Color online: Ground state phase diagram at $\mu=-0.2t$ and $\mid U\mid = 4t$ 
as determined based on the indicators, (i) $x$-component of average superconducting phase correlation 
($\phi^{xx}$), (ii) average magnetization ($m$) and (iii) inverse of DOS at the 
Fermi level ($1/N(0)$) (see text).}
\end{center}
\end{figure}

\vspace{0.5cm}

\noindent{\textit{Variational ground state:-}}

Even for a selected interaction strength the ground state parameter space is huge in terms of $\mu-h$. In order to get a 
handle on this parameter space we first carry out a mean field variational calculation for the ground state. In 
the spirit of MFT we stripe the pairing (auxiliary) field $\Delta_{ij}$ off all fluctuations such that 
$\mid \Delta \mid$ is now a real number. For a selected $\mu-h$ cross section we fix the 
relative phase between the $x$- and $y$-components of the superconducting phase 
$\phi^{rel}=\phi^{y}-\phi^{x}$, such that $\phi^{rel}$ can take discrete values as, 
$0, \pi/4, \pi/2...\pi$ etc. For each such choice of the relative phase, $\mid \Delta \mid$ is optimized so 
as to obtain the lowest energy configuration for the selected $\mu-h$ cross section. The process is carried 
out for different $\mu-h$ cross sections such that one obtains the optimized $\mid \Delta \mid$ and 
$\phi^{rel}$ corresponding to the minimum energy configuration. The optimized $\Delta \in \{\mid \Delta \mid, \phi^{rel}\}$ 
configurations are then used to compute the magnetization of the system for the corresponding 
$\mu-h$ cross section. The ground state phase diagram is mapped out based on these quantities in the $\mu-h$ plane.

\vspace{0.5cm} 

\noindent{\textit{Monte Carlo ground state:-}}
The ground state at different $\mu-h$ cross sections are verified by using Monte Carlo 
simulations and the results are found to be in excellent agreement with those obtained via the variational 
calculations. 
Within the framework of Monte Carlo protocol the system is cooled down from a random high temperature phase 
and is allowed to attain the lowest energy state at each temperature. Such unrestricted cooling allows us 
to retain the spatial fluctuations of all orders in the pairing field, which progressively dies out as the 
system approaches the ground state.  
For the Monte Carlo simulations the (a) average superconducting phase correlation ($\phi^{xx}$, $\phi^{yy}$) 
and (b) the average magnetization ($m$) are used as suitable indicators to map out the ground state.
\begin{figure*}                                                                                                       
\begin{center}                                                                                                        
\includegraphics[height=6.3cm,width=16cm,angle=0]{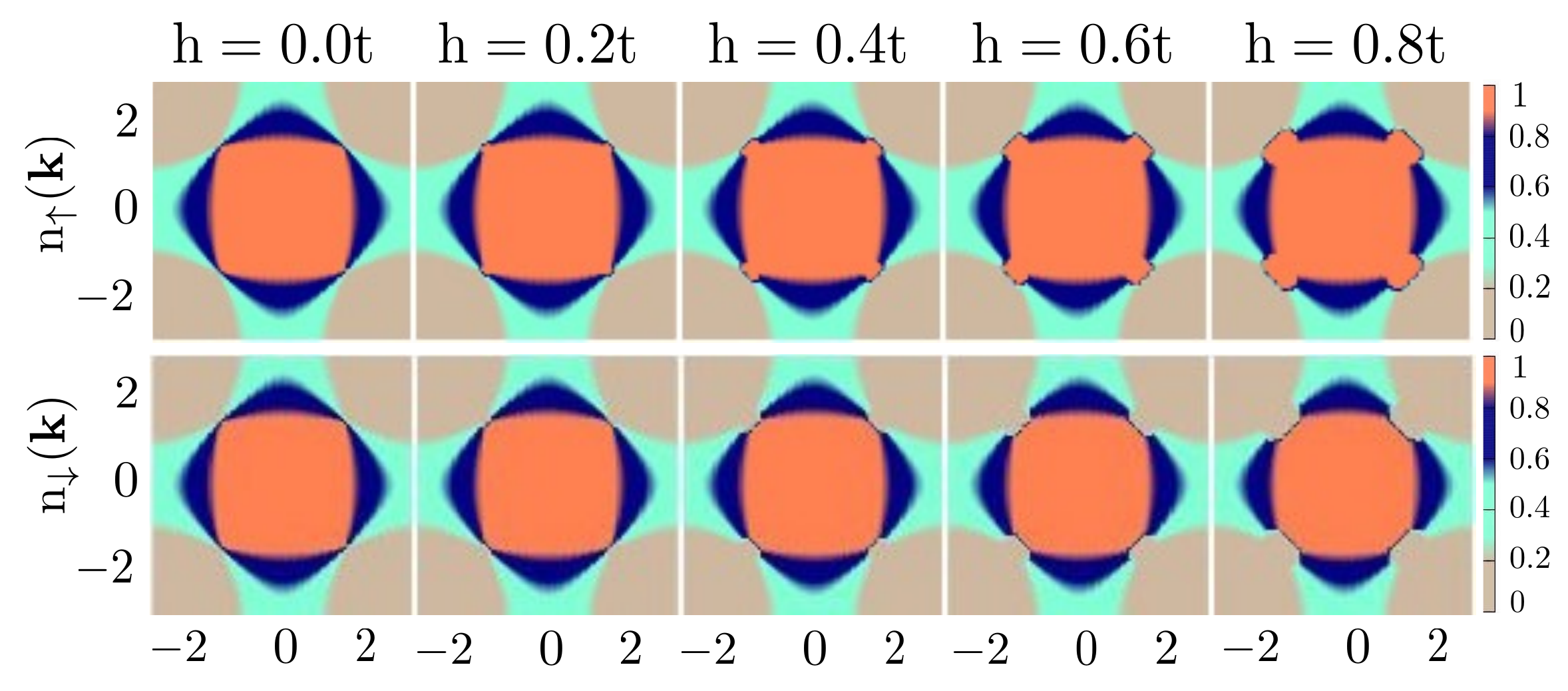}                                                             
\caption{Color online: Zeeman field dependence of spin resolved fermionic occupation                         
number ($n_{\sigma}({\bf k})$), at $T=0$. The $x$ and $y$-axes of the panels correspond to 
$k_{x}$ and $k_{y}$, respectively and the intensity of color shows the magnitude of $n_{\sigma}({\bf k})$. 
Note that the mismatch in the Fermi surface is restricted at 
the $\{\pm \pi/2, \pm \pi/2\}$ ${\bf k}$-points.}                            
\end{center}                                                                                                          
\label{gs_nocc}                                                                                                       
\end{figure*}                                                                                                         

\subsection{Parameters and indicators}
In order to study the thermal behavior of the system we select a particular cross section of the 
ground state phase diagram, such that the net chemical potential is selected to be $\mu = -0.2t$ corresponding 
to a total fermonic number density of $n \approx 0.94$. The interaction is selected to be $\mid U\mid = 4t$ 
corresponding 
to the intermediate interaction regime, which is known to be suitable for realizing $d$-wave superconductivity 
\cite{lee_rmp,dagotto2005_prl}.  Note that since the system belongs to the intermediate 
regime of interaction ($\mid U\mid \ge t$) a perturbative approach (such as MFT) breaks down at finite 
temperatures and the fluctuations dominate. One thus need to resort to non perturbative techniques 
such as the one discussed in this article to take into account the effects of fluctuations. Unlike the assumption 
of the MFT, superconductivity in this intermediate regime of interaction is lost at high temperatures via 
the loss of long range phase coherence of the superconducting pairing field rather than the suppression of the 
pairing field amplitude. The consequence are the pre-formed pairs at $T \neq 0$ leading to the pseudogap phase 
in the $d$-wave superconductors.
Our estimates based on the variational as well as Monte Carlo calculations 
suggest a $d_{x^{2}-y^{2}}$ pairing state symmetry with the relative phase between the pairing 
field components being $\phi^{rel} = \pi$, in this 
parameter regime, in agreement with the existing literature \cite{dagotto2005_prl}. 
The results presented in this paper correspond to a 
system size of $N=L \times L = 24 \times 24$, unless specified otherwise. The effect of the finite system 
size is discussed towards the end of the paper. We have analyzed our results based on the 
thermodynamic and quasiparticle indicators defined as below, 

\vspace{0.1cm}
\noindent Site resolved and averaged magnetization:- 
\begin{eqnarray}
m_{i}=n_{i\uparrow}-n_{i\downarrow} \nonumber \\
m=\frac{1}{N}\sum_{i}\langle n_{i\uparrow}-n_{i\downarrow}\rangle \nonumber
\end{eqnarray}
Average phase correlation of pairing field:- 
\begin{eqnarray}
\phi^{xx} = \frac{1}{N}\sum_{i\neq j}\langle e^{i\phi_{i}^{x}}.e^{-i\phi_{j}^{x}}\rangle \nonumber \\
\phi^{yy} = \frac{1}{N}\sum_{i\neq j}\langle e^{i\phi_{i}^{y}}.e^{-i\phi_{j}^{y}}\rangle \nonumber
\end{eqnarray}
Mixed phase correlation of pairing field:- 
\begin{eqnarray}
\phi^{xy} = \frac{1}{N}\sum_{i\neq j}\langle e^{i\phi_{i}^{x}}.e^{-i\phi_{j}^{y}}\rangle \nonumber
\end{eqnarray}
Spin resolved single particle fermionic density of states (DOS):- 
\begin{eqnarray}
N_{\uparrow}(\omega)=\frac{1}{N}\langle\sum_{i}\mid u_{n}^{i}\mid^{2}\delta(\omega-E_{n})\rangle\nonumber \\
N_{\downarrow}(\omega)=\frac{1}{N}\langle\sum_{i}\mid v_{n}^{i}\mid^{2}\delta(\omega+E_{n}) \rangle\nonumber
\end{eqnarray}
Spectral lineshapes:-
\begin{eqnarray}
A_{\sigma}({\bf k}, \omega) = -(1/\pi) Im G_{\sigma}({\bf k}, \omega)
\end{eqnarray}
Fermion occupation number:- 
\begin{eqnarray}
n_{\sigma}({\bf k}) = \langle c_{{\bf k}\sigma}^{\dagger}c_{{\bf k}\sigma}\rangle 
\end{eqnarray}
Distribution of pairing field amplitude:-
\begin{eqnarray}
P(\mid \Delta_{ij} \mid) = \langle \sum_{i\neq j}\delta(\mid\Delta \mid-\mid \Delta_{ij}\mid)\rangle
\end{eqnarray}

\noindent where, $i$ and $j$ correspond to two different sites on the lattice. $\langle \cdots \rangle$ correspond 
to thermal average and $\sigma$ is the spin label. $n_{i\sigma}$ are the number  
of the individual fermionic species, while $u_{n}^{i}$ and $v_{n}^{i}$ are Bogoliubov 
eigenfunctions corresponding to the eigenvalue $E_{n}$. $n_{\sigma}({\bf k})$ is the Fourier transform 
of the single particle Green's function. $G({\bf k}, \omega) = lim_{\delta \rightarrow 0}G({\bf k}, i\omega_{n})|_{i\omega_{n} \rightarrow \omega+i\delta}$, where $G({\bf k}, i\omega_{n})$ is the imaginary frequency transform 
of $\langle c_{\bf k}(\tau)c^{\dagger}_{\bf k}(0)\rangle$. 

\begin{figure*}                                                                                                   
\begin{center}
\includegraphics[height=7.5cm,width=15cm,angle=0]{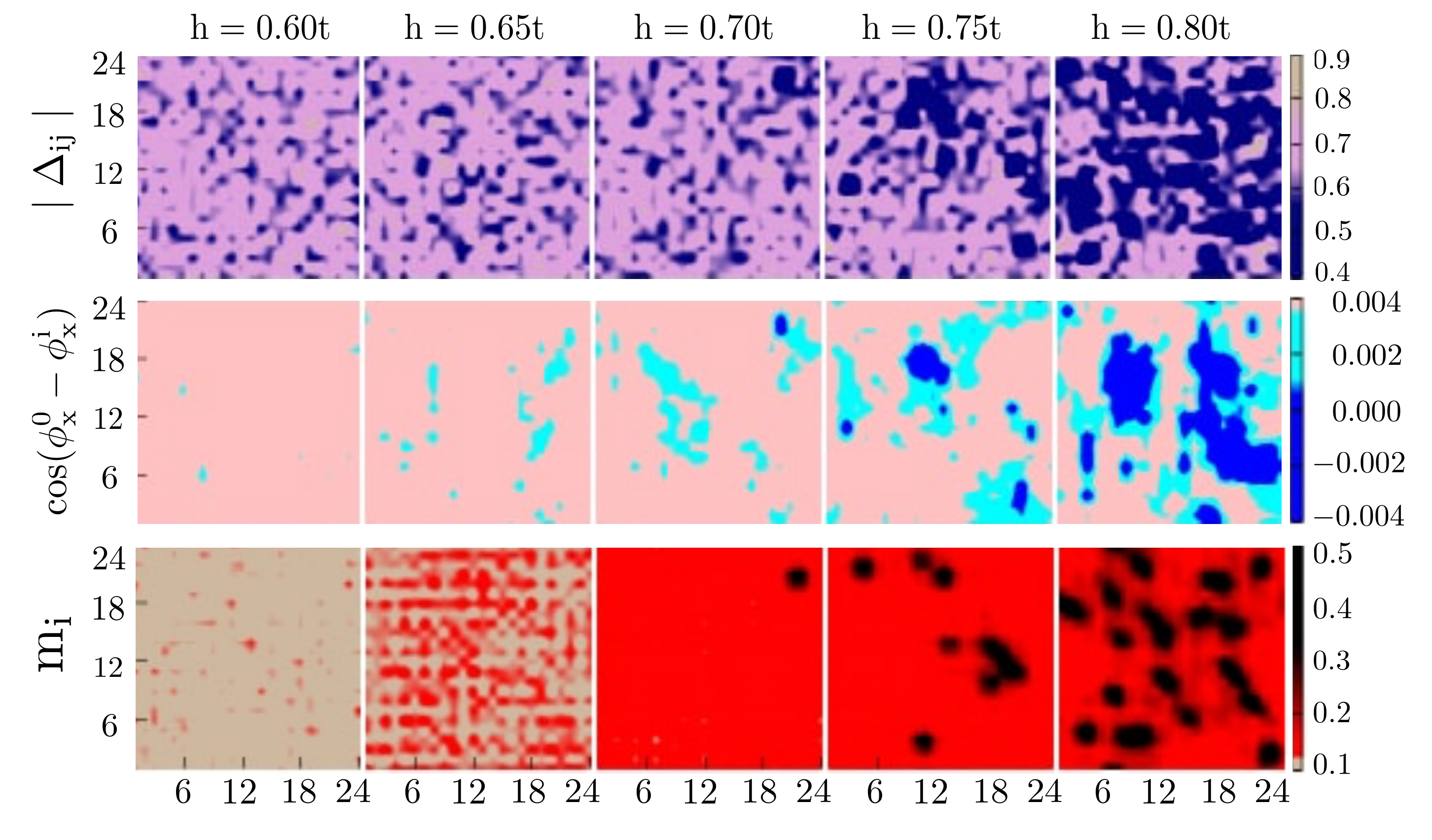}                                                    
\caption{Color online: Real space maps at $T=0$ showing the evolution of the QBP phase over a range of Zeeman field. 
The $x$ and $y$-axes of each panel correspond to the lattice indices, while the intensity of the color 
signifies the magnitude of the corresponding indicator. 
The first row corresponds to the amplitude of superconducting pairing field ($\mid \Delta_{ij} \mid$), middle
row shows the $x$-component of phase correlation of the superconducting pairing field 
($\cos(\phi_{x}^{0}-\phi_{x}^{i})$), and 
the bottom row shows the magnetization ($m_{i}$). Note that the pairing 
field undergoes depletion at isolated regions in space while the emergence of magnetization complements this 
depletion.}
\end{center}
\label{gs_map}
\end{figure*}                                                                                                     

\section{Results}

Fig. 1 constitutes one of the main results of this work, wherein 
we show the magnetization-temperature ($m-T$) phase diagram of Pauli limited 
$d$-wave superconductor. 
The highlight of this phase diagram is the $T=0$ weak magnetization regime marked as the 
quantum breached pair (QBP) phase. 
Finite temperature leads the QBP phase to undergo smooth
crossover to its classical counterpart and over a large part of the phase diagram 
the BP phase undergoes second order transition to partially polarized Fermi liquid (PPFL). 
The strong magnetization regime of the phase diagram 
belongs to the FFLO phase characterized by finite momentum superconducting pairing and first order 
thermal transition to PPFL phase. Sandwiched between the two is the unstable (phase separated) 
regime where the BP phase undergoes weak first order thermal transition to PPFL phase.  

The phase diagram shows that the QBP phase undergoes a quantum phase 
transition (QPT) to the USC phase, as a function of decreasing magnetization. 
The mechanism which drives the QPT is the Pauli paramagnetic pair breaking and 
generation of low energy gapless excitation, which hosts unpaired fermions. For a population 
imbalanced $s$-wave superconductor the gapless superconducting state is realized beyond a critical 
Zeeman field $h \ge h_{c1}$ (or population imbalance). Concomitant to this phenomena the system 
becomes unstable towards ${\bf q}=0$ superconducting pairing and the system undergoes transition to 
the FFLO phase. On the contrary, for a nodal $d$-wave superconductor, the critical field for 
paramagnetic pair breaking and generation of gapless superconducting states is different from 
the one ($h_{c2}$, say) at which the system undergoes transition to FFLO state. Over the regime 
$h_{c1} < h \le h_{c2}$ the system thus hosts a coexistent QBP phase comprising of ``gapless'' 
${\bf q} = 0$ superconducting state and non zero magnetization.
In what follows, we establish and analyze this phase diagram based on the thermodynamic and quasiparticle indicators. 

\subsection{Ground state}

We begin our discussion of the ground state by showing the energy landscape of the 
system. Given the huge parameter space we are in the energy optimization process (via variational calculation)
involves multiple 
steps which are summed up in Fig. 2 and are discussed below,
\begin{itemize}
\item[(a)]{As the first step we consider the population balanced system ($h=0.0t$) at a fixed $\mu = -0.2t$ 
and optimize the energy at different interactions $U/t$ over $\mid \Delta \mid$ and $\phi^{rel}$ (in the 
absence of fluctuations both these parameters are real numbers). The results obtained via this 
optimization is shown in Fig. 2(a) and we note that for weaker interactions the relative phase 
corresponding to the minimum energy is $\pi$ showing a $d_{x^{2}-y^{2}}$ pairing state symmetry 
of the superconducting state, in agreement with the existing literature \cite{dagotto2005_prl}. 
Increasing $U$ shifts the relative phase minima to a lower value 
such that for $7t \le U < 9t$ a $d_{x^{2}-y^{2}}+id_{xy}$ pairing state with $\phi^{rel} = 2\pi/3$ is 
stabilized. At still stronger interactions $\phi^{rel}$ shifts to even lower values. 
The results discussed in this article 
corresponds to $\mid U \mid= 4t$ and our energy minimization suggests a $d_{x^{2}-y^{2}}$ pairing state 
as the stable ground state at this interaction. For the minimized energy the corresponding 
value of $\mid \Delta\mid$ (not shown in this figure) gives the pairing field amplitude. 
The inset of Fig. 2(a) sums up the change 
in $\phi^{rel}$ w. r. t. $U/t$.}
\item[(b)]{ We next select the particular interaction $\mid U\mid =4t$, introduce population imbalance 
via the Zeeman field $h/t$ and optimize the energy over $|\Delta|(\cos({\bf q.r}))$ and 
$\phi^{rel}$, where ${\bf q}$ corresponds to the pairing momentum; for the uniform $d$-wave superconductor 
${\bf q}=0$. Note that there are now three variational parameters as $|\Delta|$, $\phi^{rel}$ and ${\bf q}$. 
This is to verify whether the pairing state undergoes a change (in terms of $\phi^{rel}$) with 
the imbalance. We show our results in Fig. 2(b) as the change in energy w. r. t. $h/t$ for selected 
$\phi^{rel}$. At all values of $h/t$ the minimum energy configuration clearly corresponds to the the relative 
phase $\phi^{rel} = \pi$, at this interaction, suggesting that the pairing state symmetry of the 
superconducting state remains unaltered with population imbalance. This narrows down the parameter 
space to $\mid U\mid = 4t$, $\mu = -0.2t$ and $\phi^{rel}=\pi$, and different $h/t$. The remaining 
task is to determine how the pairing field amplitude $\mid \Delta\mid$ varies with $h/t$, for this 
choice of the parameters. This would enable us to determine the phase boundary between the 
superconducting and non superconducting phases at the ground state. Additionally, one needs to 
keep track of the pairing momentum ${\bf q}$ so as to be able to determine the phase boundary between 
the uniform and non uniform (FFLO) superconducting phases.}
\item[(c)]{The minimization over $\mid \Delta\mid$ and ${\bf q}$ is carried out over different choices 
of spatial modulations such as,  $\Delta_{ij} \sim \mid \Delta \mid \cos(qx_{i})$, $\Delta_{ij} \sim 
\mid \Delta\mid \cos[q(x_{i}+y_{j})]$, $\Delta_{ij} \sim \mid \Delta \mid[\cos(q.x_{i})+\cos(q.y_{j})]$ 
etc. Since we are not discussing the FFLO phase in detail in this article, we show the energy landscape 
for one such choice of $\Delta_{ij}$ at different $h/t$, in Fig. 2(c). Note that for weak imbalance 
the $\mid \Delta\mid$ corresponding to the minimum energy state does not undergo significant change 
with $h/t$, larger $h/t$ leads to the development of a weak minimum at smaller $\mid \Delta \mid$ value 
indicating that population imbalance leads to suppression in the pairing field amplitude. This is 
expected because with increasing imbalance there are now lesser number of fermions which can undergo pairing.}
\item[(d)]{Finally in Fig. 2(d) we show how the pairing momentum ${\bf q} = \sqrt(q_{x}^{2}+q_{y}^{2})$ 
varies with the population imbalance. For $h \gtrsim 0.9t$ we note that the pairing momentum ${\bf q}$ picks 
up a non zero value indicating an FFLO state. Fig. 2(c) and 2(d) together gives the phase boundary between the 
uniform and FFLO superconducting states.}
\end{itemize}

The energy minimization process is carried out over different $\mu - h$ cross sections 
so as to map out the ground state phase diagram. By computing the magnetization over the optimized 
configurations one can demarcate the regimes of coexisting superconducting order and non zero magnetization.
The ground state obtained from variational calculations is reconfirmed by Monte Carlo simulation, 
wherein fluctuations are taken into account in the pairing field amplitude and phase, i. e. 
$\mid \Delta\mid \rightarrow \mid \Delta_{ij}\mid$ and $\phi^{rel}=\phi_{ij}^{y}-\phi_{ij}^{x}$, 
respectively.

We next proceed to analyze the ground state thus obtained, in terms of different indicators.
Fig. 3(a) shows the ground state phase diagram of the system in the $\mu-h$ plane, 
as determined through Monte Carlo simulation at a selected interaction strength of $\mid U \mid = 4t$.  
In the $\mu-h$ plane the system hosts four different phases as, 
(i) unmagnetized $d$-wave superconductor
(USC), (ii) coexistence, (iii) Fulde-Ferrell-Larkin-Ovchinnikov (FFLO) and (iv) partially polarized 
Fermi liquid (PPFL).
The phases are demarcated by critical magnetic fields $h_{c1}$, $h_{c2}$ and $h_{c3}$, corresponding to a second order 
 phase transition from the USC to coexistence phase at $h_{c1} \sim 0.2t$, a first order transition of coexistence 
to FFLO at $h_{c2} \sim 0.9t$ and a second order transition from FFLO to PPFL phase at $h_{c3} \sim 1.25t$, respectively. 
We note that the order of phase transition from the FFLO to PPFL phase is in 
agreement with the existing results \cite{ting2009}. The phase transition between the USC and 
coexistence phase, discussed for the first time in this paper, is of second order, 
as expected \cite{wilzek}. 

The indicators used to demarcate these phases are the $x$-component of the average phase correlation
$\phi^{xx}$ (the $y$-component of the phase correlation behaves identically) and 
the average magnetization $m$. These indicators however does not give any information regarding 
the spectral behavior of the underlying state. Consequently, based solely on these indicators a distinction 
between the superconducting state being gapped or gapless can not be made. The regime $h_{c1} < h \le h_{c2}$ is 
thus shown as the ``coexistence'' phase. While both the USC ($0 < h \le h_{c1}$) and coexistence 
($h_{c1} < h \le h_{c2}$) phases comprise of ${\bf q}=0$ $d$-wave pairing, the distinction between the gapped 
and gapless superconductivity can be made via the single particle DOS, which we discuss next. A (nodal) gapped  
superconductor with finite magnetization would correspond to a phase separated state, while a gapless 
superconductor with finite magnetization corresponds to a QBP phase. 
While our calculations are carried out in the grand canonical ensemble we recast the phase diagram in the $n-m$-plane
in Fig. 3(b). The USC phase now collapses to the x-axis, while the coexistence, FFLO and PPFL 
phases are demarcated by $m_{c1}$ and $m_{c2}$ corresponding to the critical magnetization. 
We choose a particular cross section
of this phase diagram at $\mu=-0.2t$ ($n \approx 0.94$) to understand the physics of the system both at
the ground state as well as at the finite temperatures. We have verified that the qualitative behavior of
the system is independent of the choice of $\mu$ (or $n$), except for very small filling. 
The Zeeman field regime being probed is $h=[0:1.4]$ 
where $h=[0:0.9]$ correspond to the ``notional'' weak imbalance regime, where the superconducting 
state has ${\bf q}=0$ pairing. 

In order to characterize the underlying superconducting state we next show the spin 
resolved single particle DOS at different
Zeeman field, in Fig. 4. In the absence of any population imbalance the DOS corresponding 
to the two spin species are identical and centered around the Fermi level ($\omega = 0$), 
giving rise to a nodal gap. The $d_{x^{2}-y^{2}}$
character of the pairing field is evident with $N(\omega) \propto \omega$ as 
$\omega \rightarrow 0$. Increasing imbalance shifts the spin resolved DOS away from $\omega=0$ with 
them being now centered around the shifted Fermi level $\omega = \pm h$. 
However, an interesting behavior is observed at $\omega=0$ where there is now an 
accumulation of finite spectral weight leading to a {\it gapless superconducting state}. 
The behavior of the DOS is in dramatic contrast with that of imbalanced $s$-wave superconductors where 
even though the DOS corresponding to the different spin species are centered around the 
shifted Fermi levels, the spectrum remains gapped and there are no low energy 
states at $\omega=0$,  unless the system hits the FFLO phase \cite{mpk2016}. 
The coexistence phase shown in Fig. 3 thus hosts a gapless 
${\bf q}=0$ superconducting phase along with a nonzero magnetization, characteristic 
to the QBP phase. 
The transition from the gapped to gapless superconducting phases is the 
signature of the quantum phase transition between the USC and QBP phases. 

The information obtained from the thermodynamic and quasiparticle signatures 
is summed up in Fig. 5 which shows the ground state phase diagram at $\mu = -0.2t$ and $\mid U\mid =4t$. The phase 
diagram shows only the USC and QBP phases and for the time being we ignore the ${\bf q}\neq 0$ 
pairing state. There are three indicators based on which the phases are demarcated, along with 
the $x$-component of the average phase correlation ($\phi^{xx}$) and the average magnetization 
($m$) we now show the inverse of the DOS at the Fermi level ($1/N(0)$) which gives a measure 
of the energy gap ($E_{g} \sim 1/N(0)$). As $N(0) \rightarrow \infty$, $E_{g} \rightarrow 0$, i. e. a large 
spectral weight at the Fermi level correspond to a gapless superconducting state. 
 Based on this indicator we show that for $h \le 0.2t$ 
the system in a gapped superconductor with zero magnetization corresponding to the USC phase. 
In the regime $0.2t < h \le 0.85t$,  $1/N(0)$ vanishes indicating a gapless phase, simultaneous 
non zero $\phi^{xx}$ and $m$ qualifies the phase to be a QBP. For $h \gtrsim 0.9t$, $\phi^{xx}$ collapses 
to zero while $m$ remains finite, suggesting a non superconducting PPFL phase.

Having mapped out the ground state phase diagram we next move on to a deeper 
analysis of these phases. We begin with 
the spin resolved momentum occupation number ($n_{\sigma}({\bf k})$) as a function of increasing 
Zeeman field, shown in Fig. 6. The momentum occupation number, which is one of the standard indicators 
to map out the underlying Fermi surface of the superconducting state, is the Fourier transform of the single 
particle Green's function \cite{rajdeep}. 
In the regime of balanced ($h=0t$) or very weakly 
imbalanced population the momentum occupation number corresponding to the two spin                                     
species are identical, i. e. they exhibit the usual Fermi type distribution corresponding to 
the $d$-wave superconducting state. 
Increasing imbalance leads to mismatch in this distribution and accumulation                                        
(or depletion) of weight at isolated ${\bf k}$-points, along the nodal directions ($\pm \pi/2,\pm \pi/2$), 
of the Brillouin zone.
The $n_{\sigma}({\bf k})$ is thus an important quantity which gives evidence of the change in the 
``notional'' Fermi surface topology with population imbalance. In that spirit, $n_{\sigma}({\bf k})$ 
indicates a Lifshitz like transition between the USC to QBP phases. Along the nodal directions the 
Fermi surfaces changes from being identical to disparate with increasing imbalance across this 
transition, which can be considered as a signature of the Lifshitz like transition \cite{wilzek,lifshitz}.

Fig. 7 shows the real space signatures of the QBP phase as the spatial 
maps for a typical Monte Carlo 
configuration of superconducting pairing field and magnetization 
at selected Zeeman field cross sections. 
The rows in the figure correspond to three different 
quantities viz. (i) amplitude of superconducting pairing field ($\mid \Delta_{ij}\mid$), (ii) correlation 
of the $x$-component of superconducting phase between a reference site and other sites 
of the lattice ($\cos(\phi_{0}^{x}-\phi_{i}^{x})$) and (iii) site resolved magnetization ($m_{i}$). 
At the lowest field shown in the figure the superconducting state is robust with 
nearly uniform pairing field amplitude and (quasi) long range phase coherence. 
On the other hand, magnetization is very weak ($\sim 0.1$) in this regime.
With increasing Zeeman field isolated islands of {\it depleted} 
superconductivity emerges in the system, as characterized by regions with suppressed pairing 
field amplitude and phase correlation. The islands grow in size and begin to 
merge with each other as the field is increased further. 
Increase in spatial inhomogeneity promotes spectral weight accumulation at the Fermi level 
i. e. a gapless superconducting state. Interestingly, at these 
regions of depleted superconductivity the magnetization begins to gain weight 
as shown in the bottom row of Fig. 7. 
The snapshots reveal that over a regime of Zeeman field gapless superconductivity
coexists with nonzero magnetization, characteristic to the QBP phase.

After establishing the ground state behavior of the system we now focus on the thermal 
evolution of the various phases. At the onset,  we note that the thermal transition discussed 
in this paper are basically Berezinsky-Kosterlitz-Thouless (BKT) transitions, corresponding to algebraic 
decay of quasi long range order in two-dimensions.

\subsection{Finite temperature}

We show thermal evolution of the global indicators viz. average superconducting phase correlations 
and average magnetization in Fig. 8. In panels (a)-(c) we show the field dependence 
of the average phase correlation components $\phi^{xx}$, $\phi^{yy}$ 
and $\phi^{xy}$ corresponding to the $d$-wave superconducting pairing field, across the
 USC and QBP regime, as the system evolves in temperature. Note that 
the superconducting state is fairly robust in the field regime under consideration, 
with a nonzero (indicated by the point of inflection of the curves) $T_{c}$. 
The system loses the (quasi) long range phase coherence 
at $T_{c}$, which sets the thermal transition scale of the system. In Fig. 8(d) we present the thermal 
evolution of average magnetization at different Zeeman field. Increasing temperature leads to thermal
  pair breaking and gives rise to unpaired fermions, consequently, magnetization is
larger at high temperatures. For $h \ge 0.2t$ the system is in QBP phase 
as indicated by non zero magnetization at $T=0$. 

\begin{figure}                                                                             
\begin{center}
\includegraphics[height=8.5cm,width=12.0cm,angle=0]{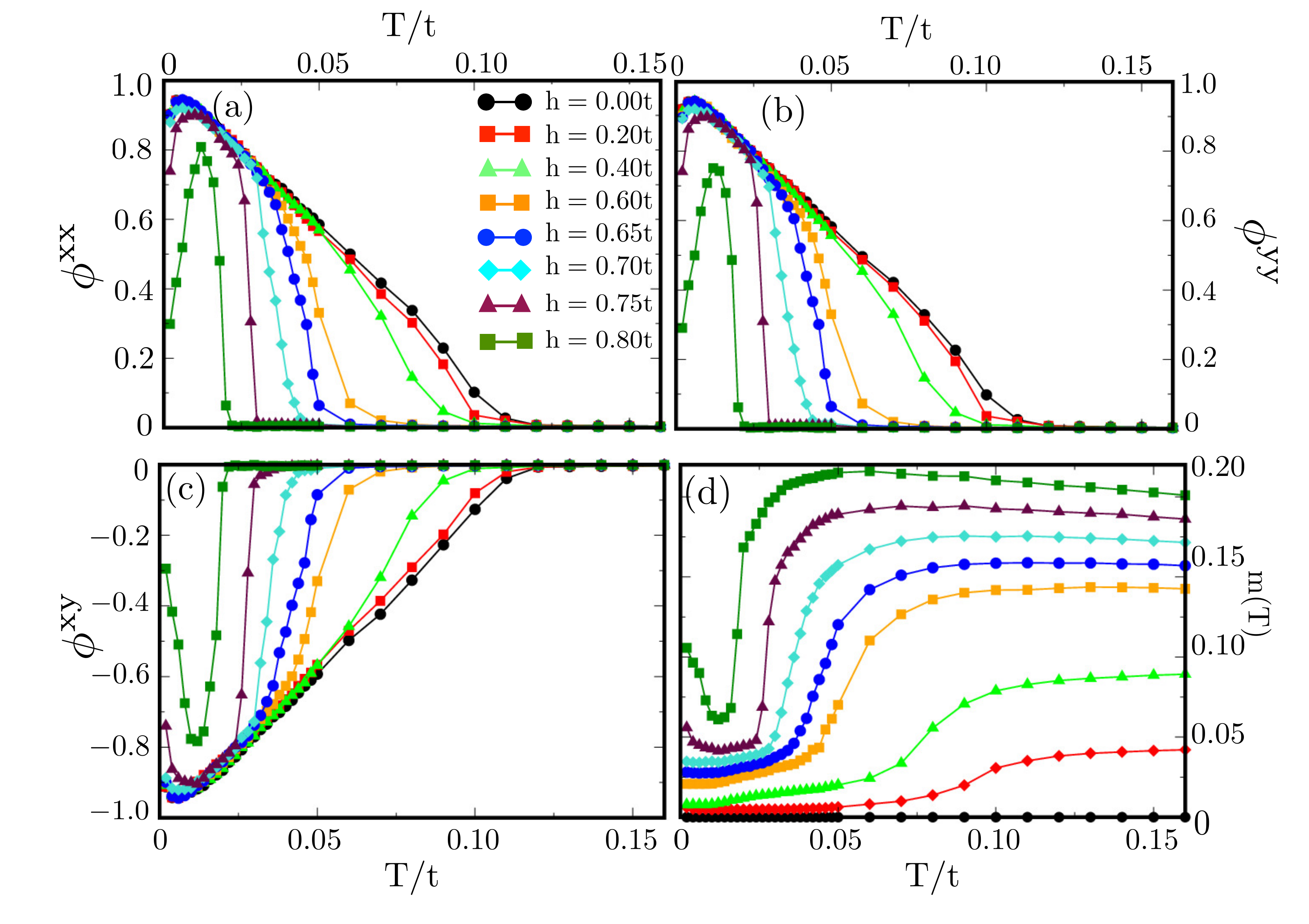}                                                    
\caption{Color online: Thermal evolution of average phase correlation (a) $\phi^{xx}$, 
(b) $\phi^{yy}$, (c) $\phi^{xy}$ and (d) average magnetization ($m$), 
with Zeeman field. The point of inflection of the curves correspond to the $T_{c}$.}
\end{center}
\label{finite_global}
\end{figure}                                                                                                     
\begin{figure*}
\begin{center}
\includegraphics[height=8.2cm,width=16.8cm,angle=0]{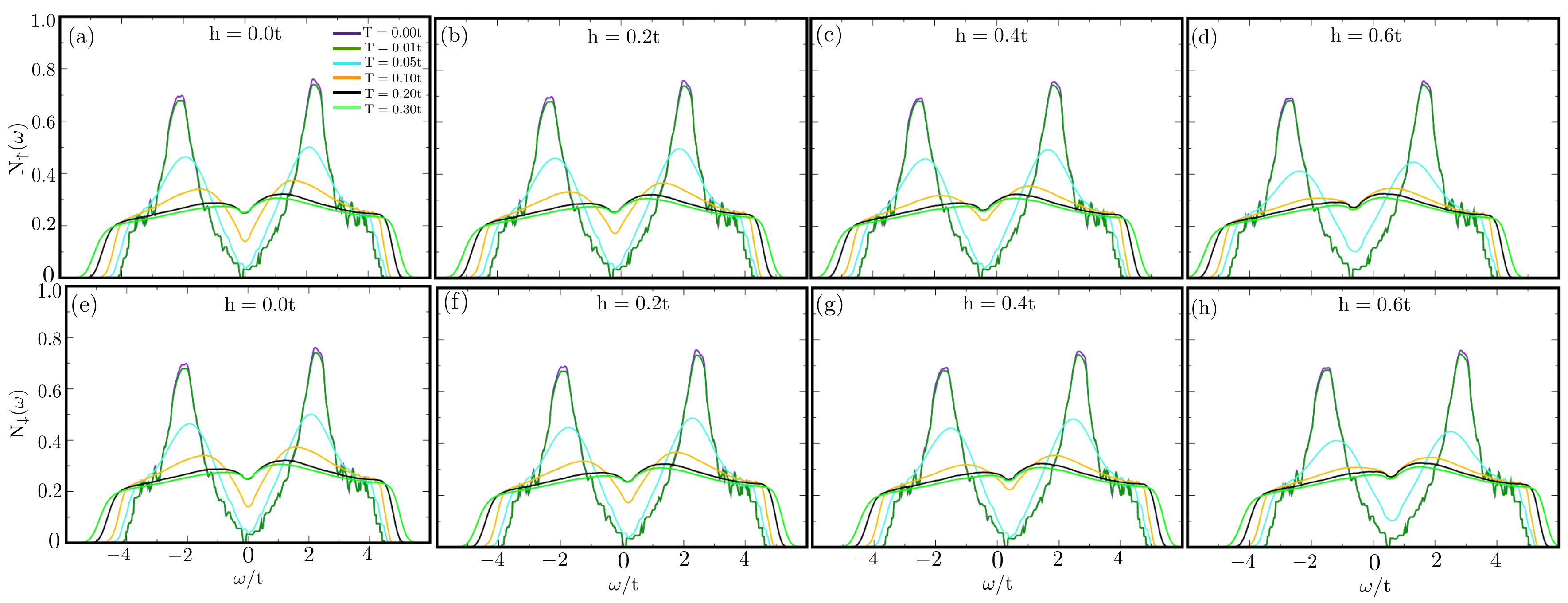}
\caption{Color online: Temperature dependence of the single particle DOS at the Fermi level
corresponding to the the up-spin (a)-(d) and down-spin (e)-(g) fermion species,
at selected Zeeman fields.}
\end{center}
\label{finite_dos}
\end{figure*}

The spin resolved DOS at different temperatures, at selected Zeeman fields are presented next in 
Fig. 9. For $h>0$ the DOS at the ground state exhibits nodal gap at the shifted 
Fermi levels, increasing temperature leads to piling up of spectral weight and progressive closure 
of these gaps. Simultaneously, the coherence peaks at the gap edges flatten out via large transfer of 
spectral weight away from the (shifted) Fermi levels, indicating the loss of (quasi) long range phase 
coherence. While the depletion of spectral weight at the shifted Fermi level is observed even at high 
temperatures owing to the correlation effects, we note that beyond a temperature, $T_{pg}$, say, the 
behavior of the gap closure becomes non monotonic. The temperature range $T_{c} < T < T_{pg}$ corresponds 
to the regime where short range superconducting pair correlations survive in the system even after the loss of 
(quasi) long range phase coherence and gives rise to the pseudogap phase. The figure demonstrates 
how the pseudogap scale undergoes progressive suppression with increasing imbalance in population. 

Thermal evolution of the momentum resolved spectral lineshapes are presented
next in Fig. 10, for a selected Zeeman field of $h=0.6t$, which corresponds to the 
QBP phase at the ground state. The figure shows the lineshape
$A_{\uparrow}({\bf k}, \omega)$ for the momentum trajectory $(0, 0) \rightarrow (\pi, 0) \rightarrow (\pi, \pi)
\rightarrow (0, 0)$, across the Brillouin zone. We note that at the lowest temperature prominent 
gap opens up along $(\pi/2, 0) \rightarrow (\pi, 0) \rightarrow (\pi, \pi/2)$. Progressive increase 
in temperature suppresses the gap and finally leads to its closure at $T \ge 1.25T_{c}$, 
where $T_{c} \sim 0.08t$. The behavior of 
$A_{\downarrow}({\bf k}, \omega)$ (not shown here) is similar to $A_{\uparrow}({\bf k}, \omega)$. Overall, 
in the nodal direction $(\pm \pi/2, \pm \pi/2)$ the lineshape is characterized by a single peak which 
is roughly immune to thermal evolution, while there is opening of gap in the antinodal 
direction which undergoes progressive 
closure with temperature. 

The real space signature is demonstrated next in Fig. 11
via the spatial maps. Once again we use (i) $\mid \Delta_{ij}\mid$, (ii) $\cos(\phi_{0}^{x}-\phi_{i}^{x})$, 
and (iii) $m_{i}$,  as our indicators and track them as they evolve in temperature at $h=0.6t$. 
Apart from the complementary spatial realization of superconducting pairing field and 
magnetization characteristic to the QBP phase at the lowest temperature, we note enhancement in 
spatial fragmentation of $\mid \Delta_{ij}\mid$ with increasing temperature. 
The observation is validated by the phase coherence maps which demonstrates the increasing inhomogeneity 
and thus the loss of (quasi) long range phase coherence due to thermal fluctuations. The fragmentation and loss of 
phase coherence of superconducting pairing field is accompanied by emergence of regions of large 
magnetization, a signature of the ``finite temperature BP phase'' \cite{mpk2016}. This regime of phase uncorrelated 
superconducting islands (at $T > T_{c}$) is the visual 
realization of the pseudogap phase, and can be thought to be a system of phase uncorrelated Josephson 
junctions. As the isolated islands of superconducting pairing field 
progressively shrinks with temperature the system loses its superconducting order and 
undergoes transition to the PPFL phase.
\begin{figure*}
\begin{center}
\includegraphics[height=7cm,width=16cm,angle=0]{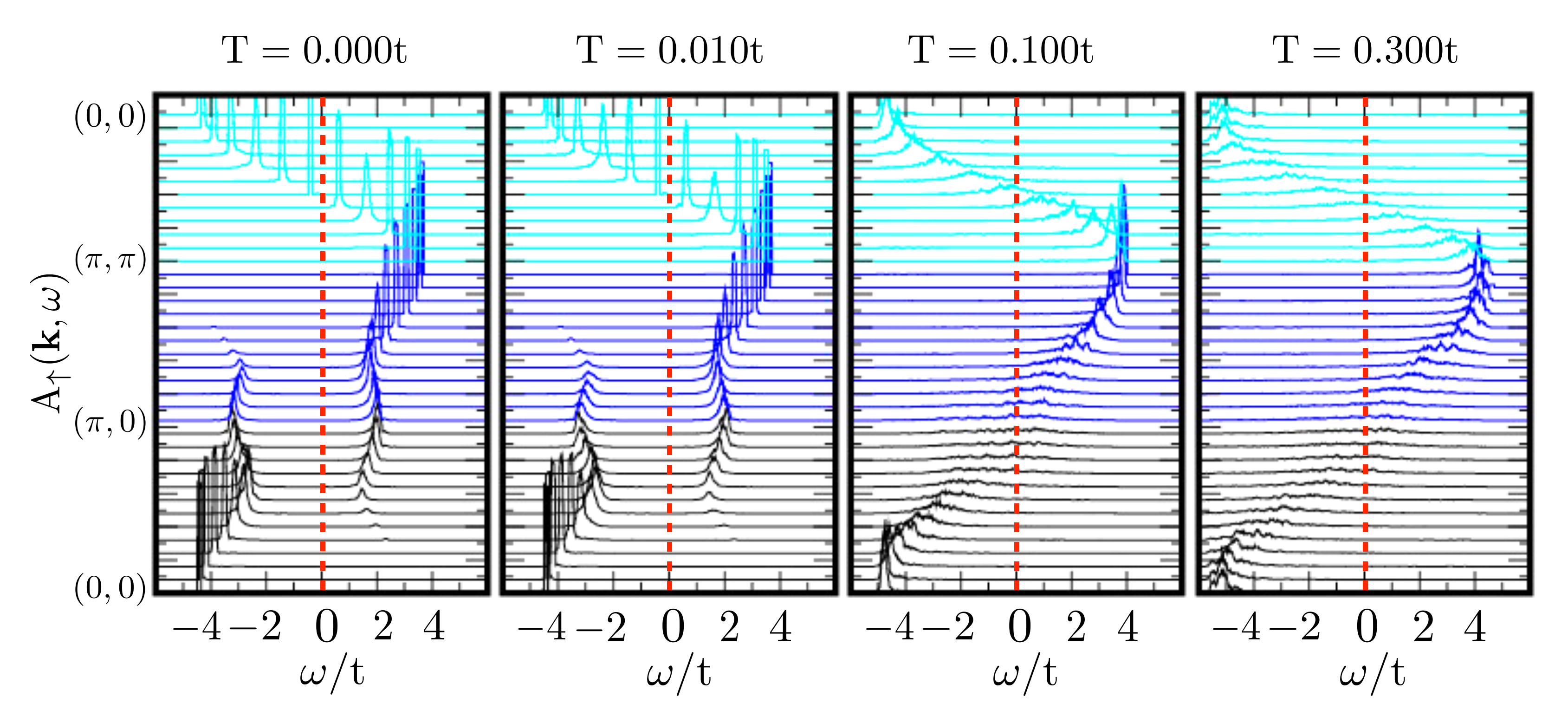}
\caption{Color online: Thermal evolution of the momentum resolved spectral line shape 
(A$_{\uparrow}({\bf k}, \omega)$) across the $(0, 0) \rightarrow (\pi, 0) \rightarrow (\pi, \pi) \rightarrow (0, 0)$ 
trajectory in the Brillouin zone, at $h=0.6t$.}
\end{center}
\label{finite_spectral}
\end{figure*}
\begin{figure*}
\begin{center}
\includegraphics[height=7.5cm,width=15cm,angle=0]{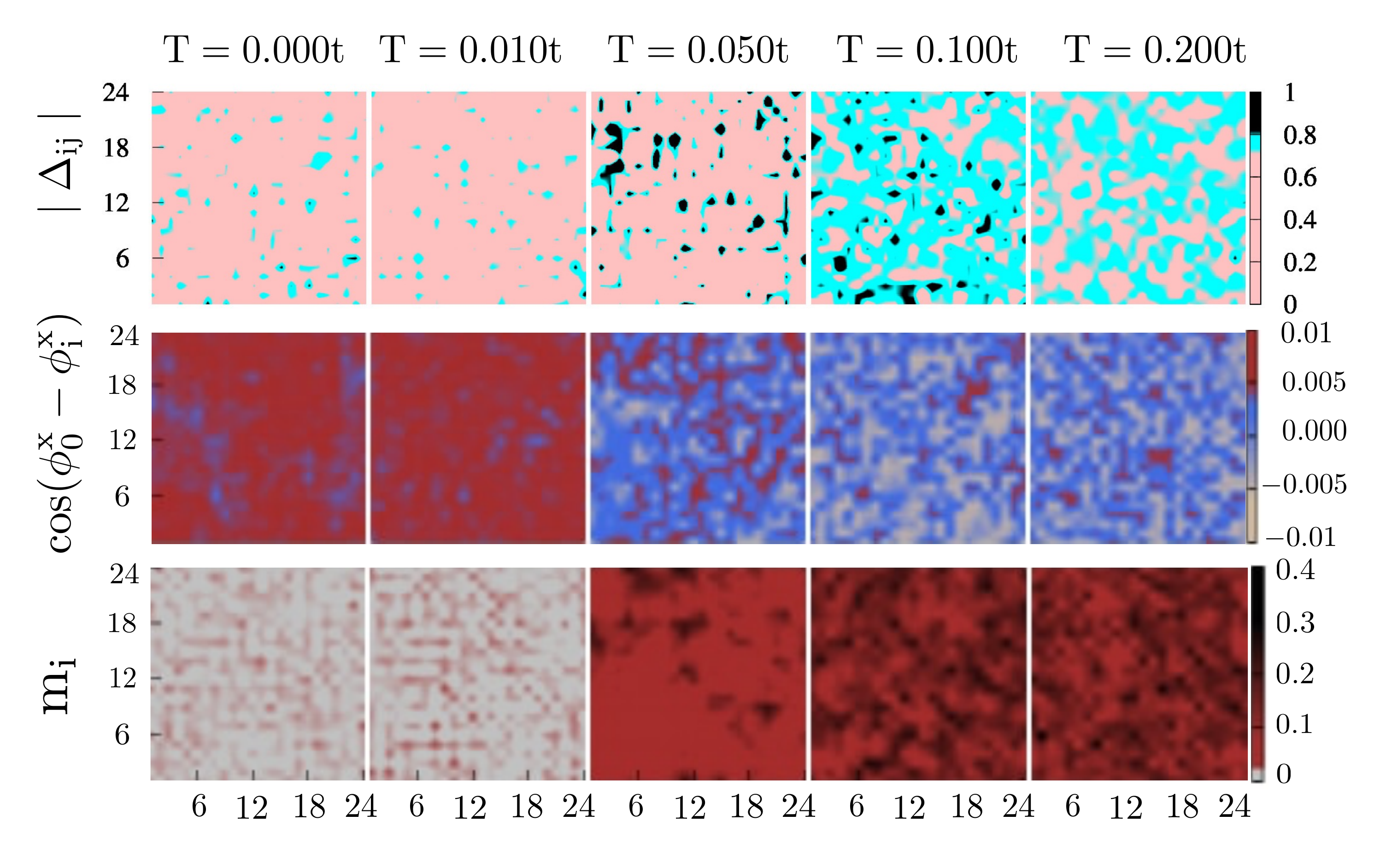}
\caption{Color online: Real space maps corresponding to the (a) pairing field amplitude
$\mid \Delta_{ij}\mid$, (b) pairing field phase coherence $\cos(\phi_{0}^{x}-\phi_{i}^{x})$ and
(c) magnetization $m_{i}$, as they evolve in temperature, at a selected Zeeman field of $h=0.6t$.
Increase in temperature leads to loss of phase coherence and eventual spatial fragmentation of the 
superconducting state.}
\end{center}
\label{finite_maps}
\end{figure*}

Note that both the QBP and the finite temperature pseudogap phases 
are characterized by spatially inhomogeneous superconducting state and non zero spectral 
weight at the Fermi level of the quasiparticle spectra. The key distinction between these two 
phases is the presence (or absence) of (quasi) long range phase coherence of the superconducting 
pairing field (i. e. the superconducting phase stiffness). While the QBP phase is characterized by a 
finite superconducting phase stiffness, 
i. e. a quasi long range phase coherence between the fragmented superconducting islands, the said 
phase coherence is lost in the finite temperature pseudogap phase, indicating the loss of global 
superconducting order. Both the USC and QBP phases undergo smooth cross over to the finite temperature 
BP phase and then undergoes transition to the pseudogap phase with increasing temperature.

We now sum up the information gathered based on our analysis of thermodynamic and quasiparticle 
indicators and revise the thermal phase diagram shown in Fig. 1. The revised phase diagram 
in the $h-T$ plane is presented in Fig. 12. 
Along with the thermodynamic phases discussed 
earlier, the figure shows the pseudogap phase over a regime of temperature $T_{c} < T < T_{pg}$, 
and $T_{pg}$ undergoes progressive suppression with 
Zeeman field. The high field regime correspond to the FFLO phase and as shown in the phase 
diagram, is characterized by large suppression in $T_{c}$, which makes its experimental 
realization, non trivial. 
Over a large part of the $h-T$ plane the system loses its global superconducting 
order via a second order thermal phase transition. The high field low temperature regime hosting the 
FFLO phase undergoes a first order thermal phase transition to lose its superconducting order. The order 
of these phase transitions are in agreement with the experimental observations of Pauli limited superconductors 
such as CeCoIn$_{5}$ \cite{matsuda2006}.

Next, we take two temperature cross sections of this phase diagram to highlight 
the quasiparticle behavior as the system transits from BP to pseudogap to PPFL 
regimes, in Fig. 13.
In panel (a) we show the up-spin DOS at $T=0.05t$ as a function of    
increasing Zeeman field. For $h \lesssim 0.6t$ the prominent coherence 
peaks signify (quasi) long range phase coherence while the system 
is in the BP phase. At $h \sim 0.7t$ the system is at the verge of transition 
to the pseudogap phase as suggested by a large accumulation of spectral weight 
at the shifted Fermi level. The coherence peaks smear out considerably and 
transfers large spectral weight away from the shifted Fermi level. The DOS 
at $h=0.8t$ and $0.9t$ are representative of the pseudogap regime. Note that the behavior of 
the DOS in this regime is different from that observed at $h=0.7t$, owing to the 
finite temperature short range FFLO fluctuations at higher Zeeman fields. $h=t$ corresponds 
to the PPFL phase at $T=0.05t$ and we find the DOS to be akin to the free electron tight binding 
spectra of a square lattice. All superconducting correlations have died out at this 
field. 

Panel (b) shows the distribution of the superconducting pairing field amplitude 
across the $T-h$ plane mentioned in panel (a). We note that 
for weak Zeeman fields the distribution remains unchanged from the one observed 
for a $d$-wave superconductor, at $h=0t$. The mean amplitude of the pairing field 
also remains roughly constant. Increasing field shifts most of the weight to 
low amplitude $\mid \Delta_{ij} \mid$s, indicating suppression of the pairing by 
Zeeman field. The distribution shows that as expected the FFLO phase comprises of superconducting 
pairing with suppressed amplitudes.

A high temperature scan at $T=0.15t > T_{c0}$ is shown in panel (c), where 
$T_{c0}$ correspond to the $T_{c}$ at $h=0t$. At all values of $h$ we find 
that the magnitude of the coherence peak is reduced to almost half of what was
observed at $T=0.05t$, as there are now only short range pair correlations 
surviving in the system {\it without} any (quasi) long range order. Consequently, the coherence 
peaks have smeared out as is expected from the pseudogap phase. The system 
transits to the PPFL phase at $h \sim 0.7t$, indicated by the complete disappearance 
of the coherence peaks. The weak depletion in the DOS at the shifted Fermi level 
as observed for $h \gtrsim 0.7t$ arises due to the strong correlation regime we are 
in, and continues to survive even at higher temperatures. The corresponding distribution 
of $\mid \Delta_{ij}\mid$ is shown in panel (d). We notice that the distribution is 
significantly broader due to thermal fluctuations and is roughly independent of the
choice of $h$ since the long range superconducting pair correlations have died
out at this temperature, as compared to its $T=0.05t$ counterpart.      
\begin{figure}
\begin{center}
\includegraphics[height=7.5cm,width=8cm,angle=0]{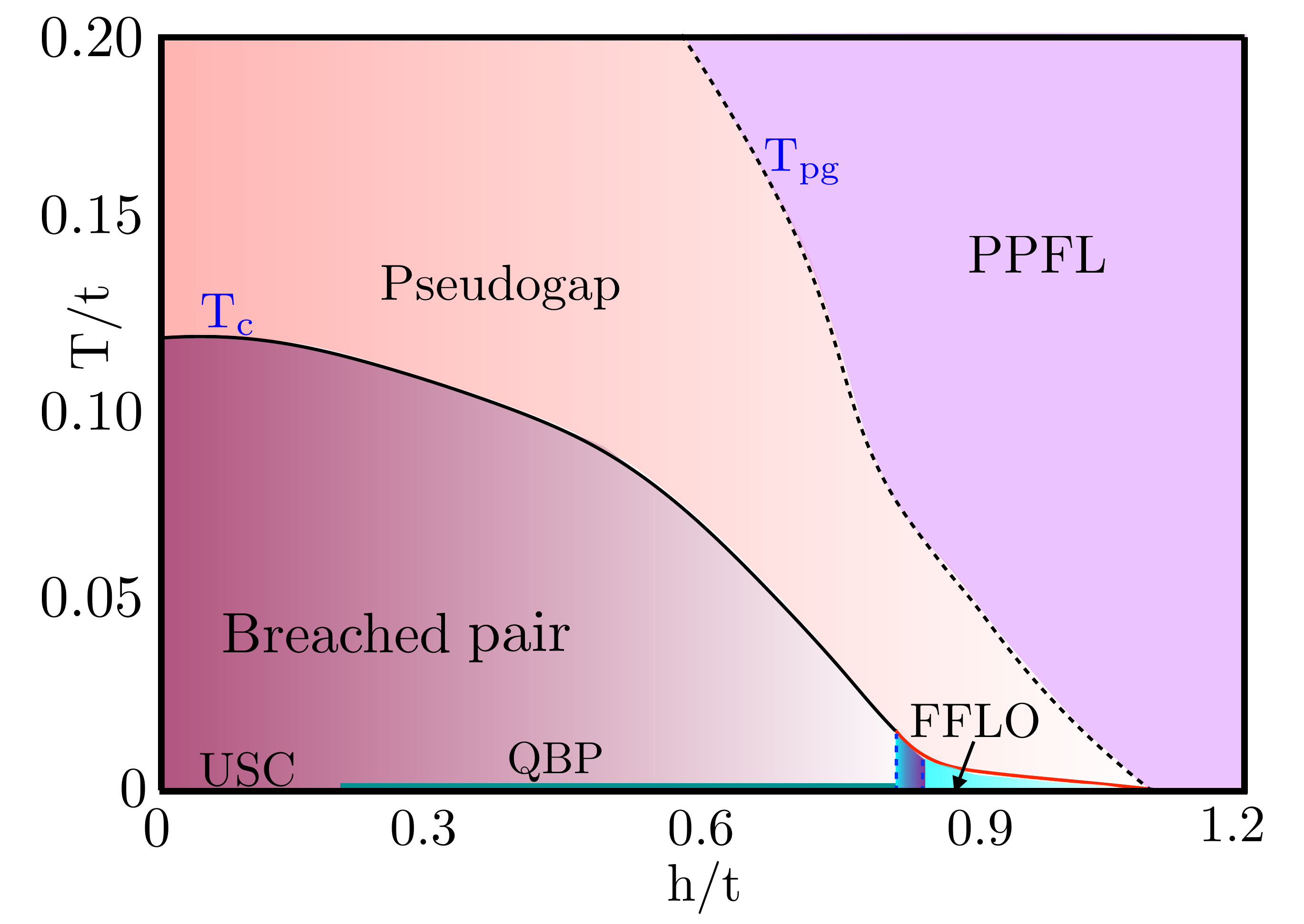}
\caption{Color online: Zeeman field-temperature ($h-T$) phase diagram of population imbalanced
$d$-wave superconductor, showing the BP, QBP,
FFLO, PPFL and pseudogap phases. The solid curve correspond to $T_{c}$, with the regime of second order
transition shown by black curve while the red curve shows the regime of first order phase transition.
The dashed curve correspond to $T_{pg}$ marking the crossover from the pseudogap to the PPFL phase.}
\end{center}
\label{finite_pd_ht}
\end{figure}

\section{Discussion and conclusion}

\subsection{Imbalance in mass}

In the earlier sections of this paper we have discussed about how a Pauli limited $d$-wave 
superconductor gives rise to a quantum breached pair phase with coexisting zero-momentum 
gapless superconducting order and nonzero magnetization. 
We demonstrated that due to the shift in the Fermi level of the individual fermion species 
there is a pile up of spectral weight at the ``unshifted'' Fermi level as suggested 
by the spatial depletion of the superconducting order. The regions of depleted superconductivity
serves as host to the unpaired fermions in the system which gives rise to nonzero 
magnetization.
\begin{figure}
\begin{center}
\includegraphics[height=9.0cm,width=12.5cm,angle=0]{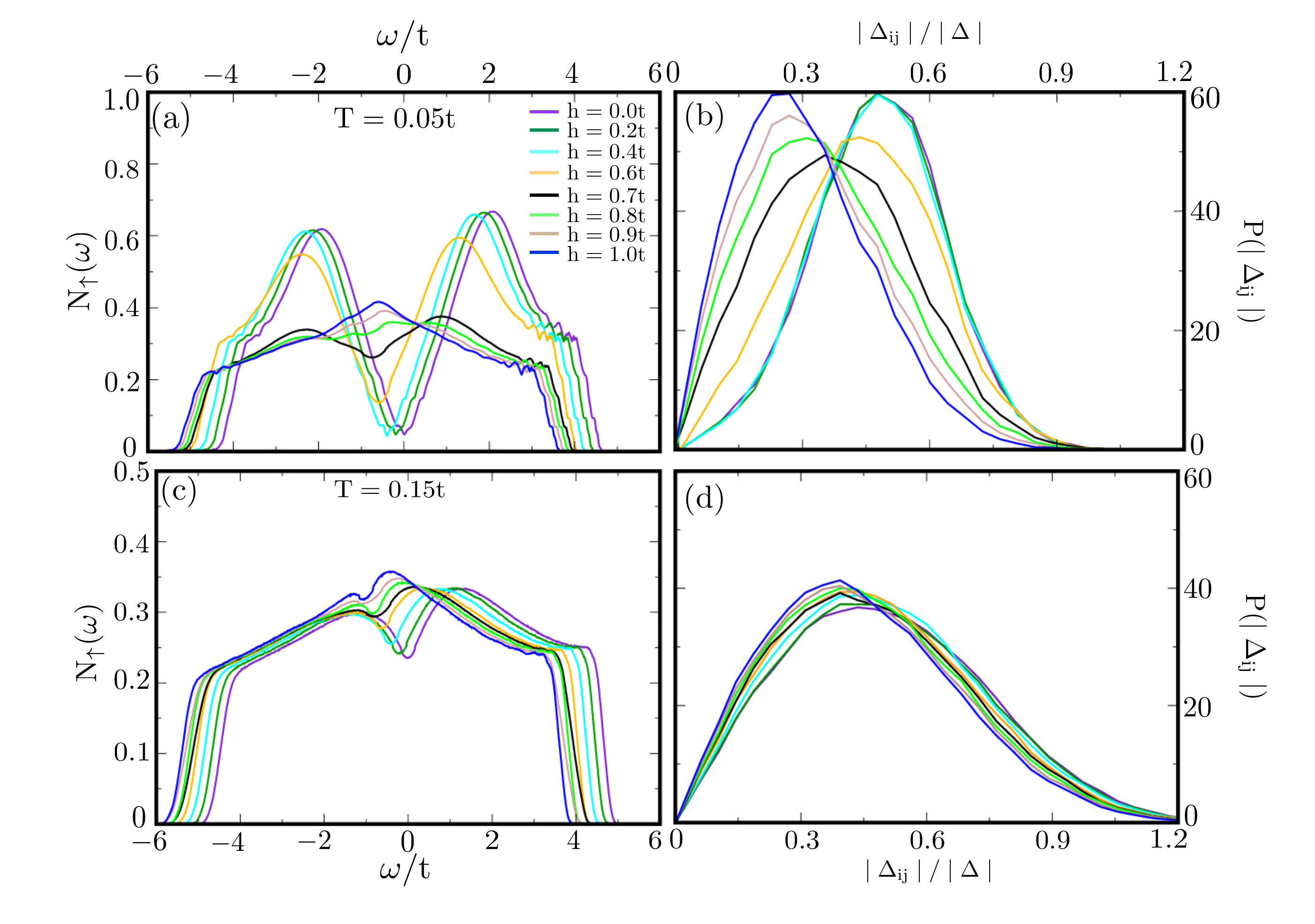}
\caption{Color online: Single particle DOS ($N_{\uparrow}(\omega)$) ((a) and (c)) and distribution 
of pairing field amplitude $P(\mid \Delta_{ij}\mid)$ ((b) and (d)) at selected temperature 
cross sections $T=0.05t$ and $T=0.15t$, respectively, for different Zeeman fields. $\mid \Delta\mid$ 
correspond to the pairing field amplitude at the ground state.}
\end{center}
\label{finite_dos_dist}
\end{figure}

In this section,  we discuss an alternative scenario where similar physical phenomena 
plays out. At this end we consider a Fermi-Fermi mixture with fermion species having 
unequal effective masses. In the context of ultracold atomic gases 
such Fermi-Fermi mixtures have already been realized as, $Li^{6}$-$K^{40}$ mixture 
\cite{tagleiber2008, voigt2009, costa2010, naik2011}, albeit 
with an isotropic $s$-wave symmetry of the pairing state. While superfluidity is 
yet to be achieved in such mixtures, the degenerate Fermi regime \cite{tagleiber2008,naik2011}, Fermi 
resonance between $Li^{6}$-$K^{40}$ atoms \cite{wille2008,costa2010,naik2011} and formation of $Li^{6}$-$K^{40}$ 
hetero molecules \cite{voigt2009} is already a reality. Moreover, other Fermi-Fermi mixtures such as, 
$Dy^{161}$, $Dy^{163}$, $Er^{167}$ etc. are expected to be experimentally realizable in 
the near future \cite{lev2012,kotochigova2013}. The theoretical efforts that have been 
put in to understand such hetero molecules are primarily concentrated on continuum 
models. Density functional theory combined with local density approximation \cite{braun2014}, 
functional renormalization group analysis \cite{drut2015}, mean field theory with Gaussian 
fluctuations \cite{stoof2010,stoof2009}, T-matrix and extended T-matrix approaches 
\cite{ohashi2013,ohashi2014,levin2009} have been employed 
to understand the physics of Fermi-Fermi mixtures. Within the purview of lattice fermion models, 
quantum Monte Carlo study on one dimensional mass imbalanced system \cite{batrouni2009}, non perturbative 
lattice Monte Carlo based analysis of two-dimensional system \cite{braun2015} and a recent static path 
approximation based Monte Carlo study \cite{mpk_mass} are some of the efforts worth 
mentioning.

In the present section we present a similar scenario with the pairing symmetry being non local, a target that can 
be achieved in experiments pertaining to ultracold atomic gases. 
In the context of solid state systems, materials with unequal masses of the fermions 
can be envisaged as different fermion species belonging to different electronic bands. 
Below we demonstrate that a quantum breached pair state and the associated phase transitions can be realized in such 
systems even {\it without} an imbalance in the fermionic populations. 

The Fermi-Fermi mixture comprising of unequal mass fermion species subjected to a 
non local interaction can be depicted by the Hamiltonian,
\begin{eqnarray}
H_{SC} & = & -\sum_{\langle ij\rangle, \alpha}t_{ij}^{\alpha}(c_{i\alpha}^{\dagger}c_{j\alpha} + h. c.) + \sum_{i \neq j}\Delta_{ij}(c_{iL}^{\dagger}c_{jH}^{\dagger}+c_{jL}^{\dagger}c_{iH}^{\dagger}) + h. c. \nonumber \\ && -\mu\sum_{i\alpha}\hat n_{i\alpha} + 4\sum_{i\neq j}\frac{\mid \Delta_{ij}\mid^{2}}{\mid U\mid}
\end{eqnarray}
\noindent where, $\alpha = L, H$ corresponds to the light and heavy fermion species, respectively.
Note that the imbalance in mass is imbibed in the hopping parameter $t_{ij}^{\alpha}$, wherein, $t_{ij}^{\alpha} \sim 1/m_{\alpha}$; 
 $m_{H}$ and $m_{L}$ correspond to the masses of the heavy and light species, 
respectively. Once again we set the hopping to be nearest neighbor and the energy scales are measured in units of 
$t^{L}$, which is set to unity.
As our tuning parameter we define the ratio between the hopping parameters corresponding 
to the fermion species as, $\eta = t_{H}/t_{L} = m_{L}/m_{H}$. Thus, $\eta=1$ corresponds to the balanced
(equal mass) limit of the system and $\eta=0$ corresponds to maximum imbalance, wherein one of the 
fermion species is completely localized.  
\begin{figure}
\begin{center}
\includegraphics[height=6.5cm,width=12.0cm,angle=0]{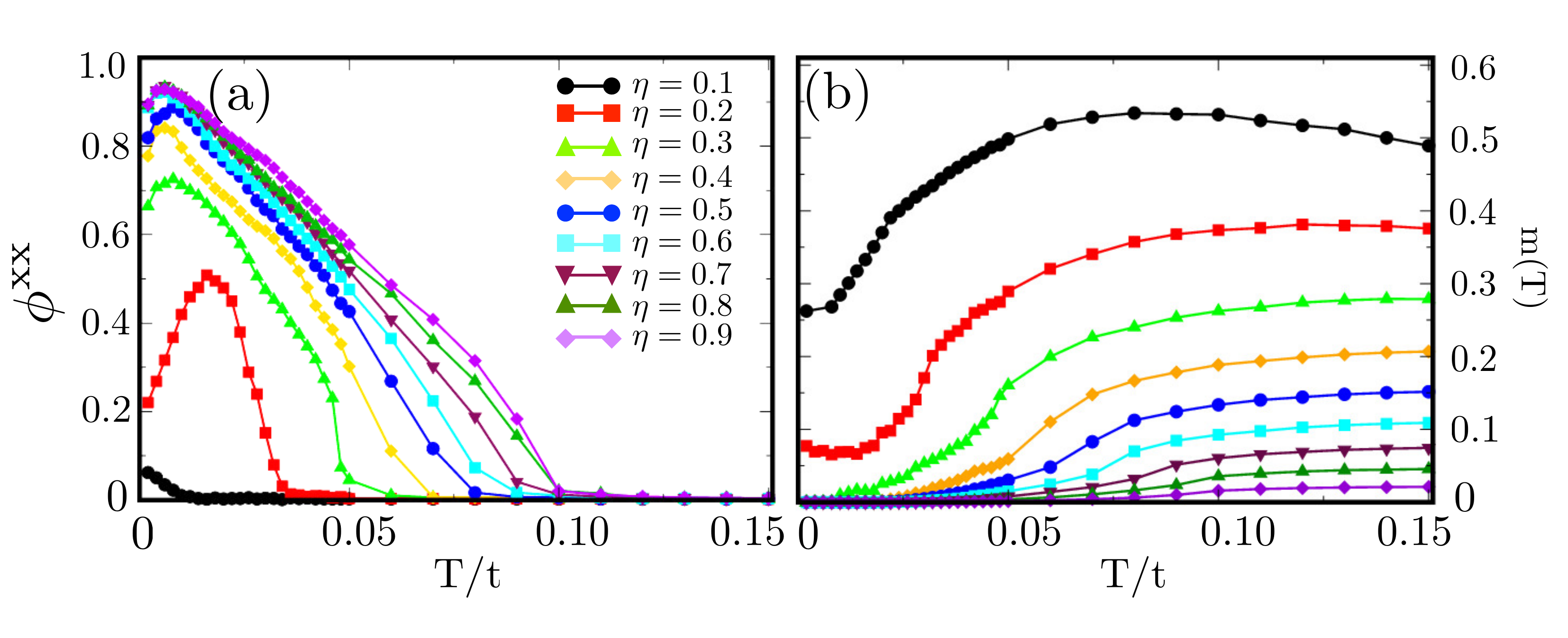}
\caption{Color online: (a) $x$-component of average pairing field 
phase coherence $\phi^{xx}$ and (b) average magnetization
($m(T)$) for $d$-wave superconductor with imbalance in the effective masses of the 
fermion species. At large imbalance  $\eta \lesssim 0.4$ the $T=0$ state shows coexisting superconducting order 
and non zero magnetization.}
\end{center}
\label{mass_global}
\end{figure}

We show the thermodynamic behavior corresponding to this system in 
Fig. 14, in terms of the $x$-component of the average phase correlation $\phi^{xx}$ and average 
magnetization $m$, at different $\eta$s. We note that though the phase correlation gets progressively suppressed with increasing mass imbalance, 
over a regime of imbalance $\eta \le 0.4$ the system hosts nonzero magnetization along with a finite phase 
correlation, at the ground state. In order to verify whether this coexistence phase is a QBP phase we have examined
the single particle DOS and have found them to be gapless.

In the limit of mass balance ($\eta=1$) the DOS corresponding to the two species 
exhibit nodal gap at the Fermi level corresponding to an uniform $d$-wave superconductor. As the imbalance 
increases one of the species gets semi localized, as a result the gain in the condensation energy through pairing
is less than that in case of the balanced mixture, i. e. not all the fermions get paired up. There is now accumulation 
of finite spectral weight at the Fermi level,  the resulting gapless spectra accommodates the excess 
(unpaired) fermions and 
give rise to a coexistence phase characteristic to the QBP. For 
weaker interaction strength, gapless superconductivity is realized for smaller imbalance ($\eta \rightarrow 1$ 
as $\mid U\mid \rightarrow 0$).

In Fig. 15 we show the composite phase diagrams for the proposed scenario of Fermi-Fermi 
mixture, both at the ground state as well as at finite temperature. The ground state 
phase diagram is shown in panel (a). In the regime of $\eta \sim 0$ the system is 
a partially polarized Fermi liquid (PPFL) with $m\neq 0$ and $\phi^{xx}=0$, with increasing $\eta$ 
the pairing field phase correlation $\phi^{xx}$ picks up weight and progressively increases 
as more fermions participate in the pairing. Simultaneously, the magnetization gets suppressed 
$m\rightarrow 0$ (due to decrease in the number of unpaired fermions) 
and eventually drops to zero at $\eta \sim 0.3$,  beyond which the system is an unmagnetized $d$-wave 
superconductor. The regime with $m \neq 0$ and
$\phi^{xx} \neq 0$ correspond to the QBP phase, in the phase diagram.
$\eta \sim 0.3$ is thus the mean field estimate of the quantum critical point for the 
quantum phase transition between the USC and QBP phases, at the parameters under consideration.
The average phase correlation ($\phi^{xx}$) continues to increase and becomes saturated 
for $\eta > 0.5$, to the magnitude expected from the balanced system.

In panel (b) we present the corresponding finite temperature behavior. The figure shows 
four distinct phases as, (i) superconductor
(BP), (ii) PG-I, (iii) PG-II and (iii) PPFL; along with three important thermal scales as, 
(i) $T_{c}$, (ii) $T_{pg}^{I}$ and (iii) $T_{pg}^{II}$. Before we characterize the phases 
individually we emphasize that unlike a system with imbalance in fermion populations, there 
are {\it two} pseudogap regimes for systems with mass imbalance, as shown in the figure. 
In one of the earlier works the author had carried out detail analysis of mass imbalanced Fermi-Fermi 
mixtures with on-site interaction \cite{mpk_mass}, such as, $Li^{6}$-$K^{40}$.
 The analysis showed that since the two fermion species are being subjected 
to different ``scaled'' temperatures, the regime over which the short range correlations survive 
in each of them are different. Consequently, while in the PG-I regime both the species are 
pseudogapped, in the PG-II regime it is only the lighter species which is pseudogapped  
while the heavier species is a partially polarized Fermi liquid. We do not expect any qualitative change from 
this picture when the pairing state symmetry is $d$-wave, as demonstrated in Fig. 15(b).
As expected, the BP regime involves finite temperature coexistence of $d$-wave superconductivity 
and non zero magnetization, while the PPFL phase is highly magnetized with vanishing superconducting 
correlations. The transitions from superconductor to PG-I, crossover from PG-I to PG-II and from 
PG-II to PPFL are marked by $T_{c}$, $T_{pg}^{I}$ and $T_{pg}^{II}$, respectively. As $\eta=1.0$
corresponds to balanced limit, the $T_{pg}^{I}$ and $T_{pg}^{II}$ scales collapse to one at this 
point. The survival of the pseudogap regimes upto $T \gg T_{c}$ ensures that signatures of mass 
imbalance in such systems can be accessed through species resolved spectroscopic experiments, 
even when the superconducting transition scales are strongly suppressed due to imbalance.  
\begin{figure}
\begin{center}
\includegraphics[height=6.5cm,width=12.0cm,angle=0]{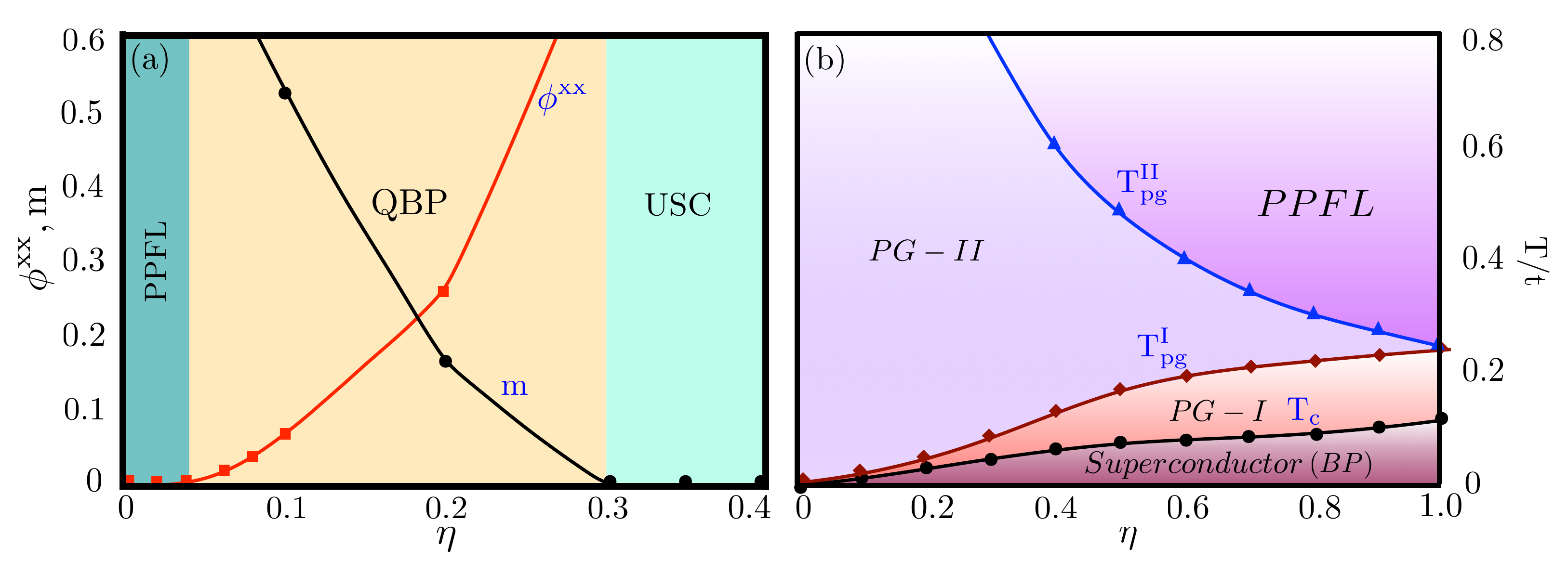}
\caption{Color online: (a) Ground state and (b) thermal phase diagram of $d$-wave superconductor 
with unequal effective masses of the fermion species. Thermal phase diagram shows species 
resolved pseudogap phases, with both the species being pseudogaped in the PG-I regime and only 
the lighter species being pseudogapped in the PG-II regime.}
\end{center}
\label{mass_pd}
\end{figure}

In conclusion, in this paper we have carried out a systematic study of Pauli limited 
 $d$-wave superconductors, within the purview of a lattice 
fermion model. Unlike the existing body of literature, we focus our analysis on the regime of 
population imbalance where it is not strong enough to give rise to the exotic FFLO phase, but at the same 
time is sufficient to give rise to Fermi surface mismatch and unpaired fermions. 
To the best of our knowledge, we for the first time demonstrate that in $d$-wave Pauli 
limited superconductors, imbalance in fermion populations give rise to a quantum phase transition 
from an uniform superconductor to a quantum breached pair phase. We give a mean field estimate 
of the quantum critical point of this transition.  
This quantum phase transition is quantified by the average magnetization of 
the system, rather than the superconducting pairing field. The merit of this work rests 
not just in establishing the quantum breached pair phase of the Pauli limited $d$-wave 
superconductors but also in accessing the thermal phases of such systems. 
While the existing literature on the Pauli limited $d$-wave superconductors seem to be
restricted to studies based on the mean field theory, we for the first time implement a non perturbative 
numerical technique which enables us to access the thermal scales accurately, owing to its inclusion of the 
spatial fluctuations of the superconducting pairing field. Apart from the population 
imbalanced $d$-wave superconductors we have 
discussed about an alternate scenario where a quantum breached pair phase can be realized, 
viz. a mass imbalanced $d$-wave superconductor. We have mapped out the ground 
state and finite temperature phase diagrams for such systems and showed that while the ground 
state does host a quantum breached pair phase, the finite temperature phase comprises of 
species selective regimes for the survival of short range pair correlations, giving rise to 
two pseudogap phases. 

While much attention has been paid to understand the FFLO physics of the Pauli limited $d$-wave 
superconductors, this is the first work which establishes the existence of a quantum breached 
pair state in these systems over a significant regime of imbalance. We have discussed several 
thermodynamic and quasiparticle indicators which should be accessible to the existing experimental 
probes.  We believe that this work is likely to open up exciting new avenues for experimental research to observe 
quantum breached pair phase in solid state materials as well as in ultracold atomic gas setups.

\section{Acknowledgements}

The author duly acknowledges the use of HPC cluster at Harish Chandra Research Institute, Prayagraj 
(Allahabad), India and Virgo HPC cluster at IIT, Madras, India. 

\appendix

\renewcommand{\theequation}{A\arabic{equation}}
\renewcommand{\thefigure}{A\arabic{figure}}
\setcounter{equation}{0}
\setcounter{figure}{0}
\section{Appendix}

\begin{figure}
\begin{center}
\includegraphics[height=6.5cm,width=12.5cm,angle=0]{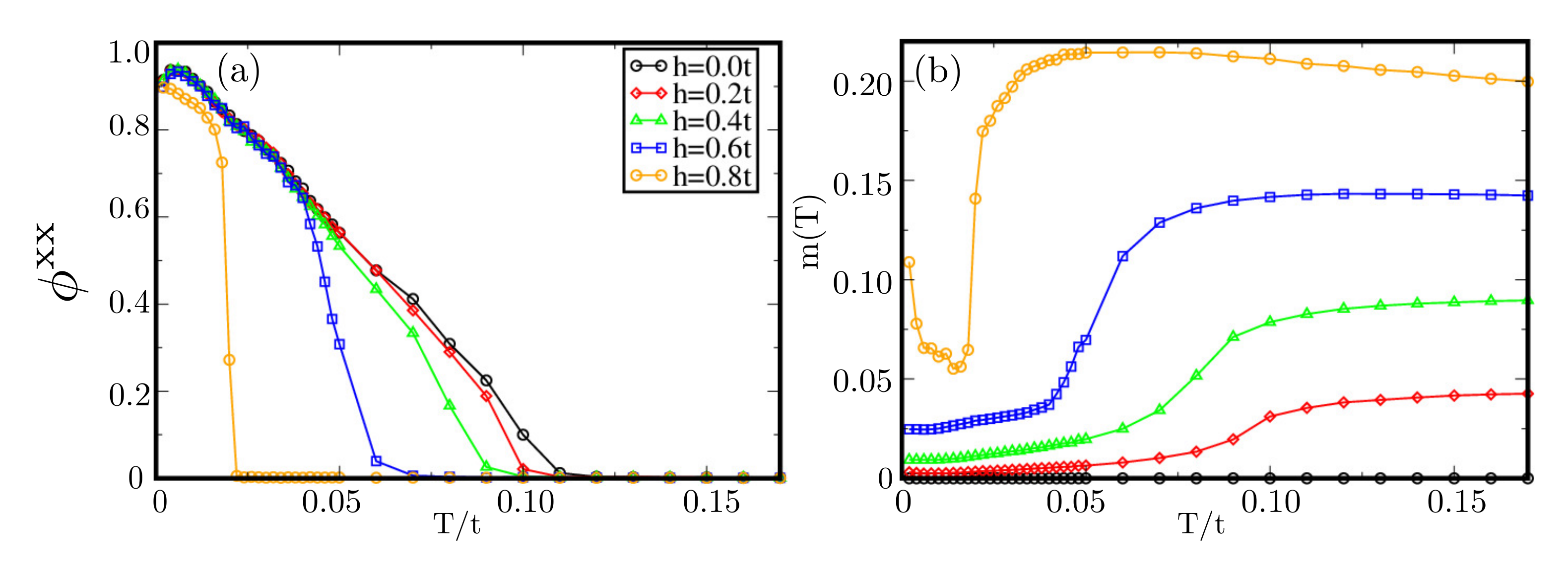}
\caption{Color online: Thermal evolution of the $x$-component of average  
superconducting phase correlation (a) $\phi^{xx}$ and (b) average magnetization
 ($m$), with Zeeman field,  for system size of $L=36$.}
\end{center}
\end{figure}

\subsection{Finite size effect}

The results presented in the main text correspond to a particular lattice size of $L=24$. 
Any lattice calculation is however likely to be plagued by finite size effect. In order to 
verify whether the quantum breached pair phase discussed in this paper is an artifact of finite 
lattice sizes we have carried out the simulations at different system sizes
(upto $L=40$). In Fig. A1 we 
show the thermal evolution of average of $x$-component of the superconducting phase correlation 
and average magnetization at $L=36$ 
for different Zeeman field. As is evident from the figure, there is indeed nonzero 
magnetization at $T=0$ for 
$h\ge h_{c1} \sim 0.2t$ suggesting that the quantum breached pair phase is robust and stable against  
finite lattice size effects.
 
\section*{References}
\bibliographystyle{iopart-num}
\bibliography{dbp}

\end{document}